\documentclass[a4paper,fleqn,usenatbib]{mnras}

\usepackage[T1]{fontenc}
\usepackage{ae,aecompl}

\usepackage{graphicx}	% Including figure files
\usepackage{amsmath}	% Advanced maths commands
\usepackage{amssymb}	% Extra maths symbols

\title[X Per: eccentric wave in the disc]{An eccentric wave in the circumstellar disc \\
of the Be/X-ray binary X Persei}

\author[Zamanov et al.]{
R. K. Zamanov,$^{1}$\thanks{E-mail: rkz@astro.bas.bg (RKZ), kstoyanov@astro.bas.bg (KAS) }
K. A. Stoyanov$^1$, 
U. Wolter$^2$,
D. Marchev$^3$,
N. A. Tomov$^1$,
\newauthor
M. F. Bode$^{4,5}$, 
Y. M. Nikolov$^{1}$,  
V. Marchev$^1$,
L. Iliev$^{1}$,
I. K. Stateva$^{1}$
\\
\\
$^{1}$ Institute of Astronomy and National Astronomical Observatory, 
       Bulgarian Academy of Sciences,  72 Tsarigradsko Shose, \\
       1784 Sofia, Bulgaria \\
$^{2}$  Hamburger Sternwarte, Universit\"at Hamburg, Gojenbergsweg 112, 21029 Hamburg, Germany  \\
$^{3}$ Department of Physics and Astronomy, Shumen University, 115 Universitetska Str., 9700 Shumen, Bulgaria \\
$^{4}$ Astrophysics Research Institute, Liverpool John Moores University, IC2, 149 Brownlow Hill, Liverpool, L3 5RF, UK \\
$^{5}$ Office of the Vice Chancellor, Botswana International University of Science and Technology, 
        Private Bag 16, Palapye, Botswana \\
}
\date{Accepted 2020 September 30. Received 2020 September 18; in original form 2020 March 16 }
\pubyear{2020?}
\begin{document}
\label{firstpage}
\pagerange{\pageref{firstpage}--\pageref{lastpage}}
\maketitle

% Abstract of the paper
\begin{abstract} 
% A series of optical spectroscopic observations 
% is presented of the Be/X-ray binary system X Per that was made over the  last four years.
We present spectroscopic observations of the Be/X-ray binary X Per
obtained during the period December 2017 - January 2020 (MJD~58095 - MJD~58865).
In December 2017 the $H\alpha$, $H\beta$, and HeI 6678 emission lines 
were symmetric with violet-to-red peak ratio 
$V/R \approx 1$.  
During the first part of the period (December 2017 - August 2018) the V/R-ratio decreased to 0.5
and the asymmetry developed simultaneously  in all three lines. 
In September 2018, 
a third component with velocity $\approx 250$~km~s$^{-1}$ appeared on the red side of the HeI line profile. 
Later this component emerged in $H\beta$, accompanied by the appearance of a red shoulder in $H\alpha$.
% Possibly, during the first half of the period  the entire disc become asymmetric and/or elliptical.
% We inperpret it as eccentric wave with eccentricity $e \approx 0.4$, outflowing velocity 
% $v_ew \sim 1$~\kms.
Assuming that it is due to an eccentric wave in the circumstellar disc, we find that 
the eccentric wave appeared first in the innermost part of the disc,
it spreads out  with outflowing velocity $v_{wave} \approx 1.1 \pm 0.2 $~km~s$^{-1}$,
and the eccentricity of the eccentric wave is $e_{wave} \approx 0.29 \pm 0.07$. 
A detailed understanding of the origin of such eccentricities would have applications 
to a wide range of systems from planetary rings to AGNs.

\end{abstract}

% Select between one and six entries from the list of approved keywords.
% Don't make up new ones.

%%%%%%%%%%%%%%%%%%%%%%%%%%%%%%%%%%%%%%%%%%%%%%%%%%

%%%%%%%%%%%%%%%%% BODY OF PAPER %%%%%%%%%%%%%%%%%%
\section{Introduction}
The relatively bright variable star 
X Persei (HD 24534)  is the optical counterpart of the X-ray source 4U 0352+309 
\citep{1972Natur.235..273B}   % Braes and Miley 1972}
and is classified in the Be/X-ray subclass of massive X-ray binary stars
\citep[e.g.][]{2007ASPC..367..477N, 2011Ap&SS.332....1R}.  %  (e.g. Negueruela 2007,  Reig 2011).
It consists of an early type Be star and a slowly spinning neutron star. 
The X-ray  data revealed a neutron star with spin period $\approx 836$~s, which  
exhibits quasi-periodic X-ray flares with a period of $\sim 7$ years \citep{2019IAUS..346..131N}.   % Nakajima.et.al 2019
The pulse period shows episodes of spin-up and spin-down  \citep{2014MNRAS.444..457A}  % (Acuner et al. 2014)
which indicates that  the neutron star  is close to a torque equilibrium 
and the long pulse period suggests a strong magnetic field 
\citep{2018PASJ...70...89Y}.     %(Yatabe et al. 2018).
\cite{2001ApJ...546..455D}        % Delgado-Mart{\'\i} et al. (2001) 
determined the orbital period $\sim$250~d,
orbital eccentricity $e$ = 0.11, semi-major axis $a$ = 2.2~a.u
and orbital inclination $i = 26^0-33^0$. 
     
During the last century, the visual brightness of X~Per has varied in the range  V= 6.8 to  6.2 mag.
% The brightness variations are accompanied by variations in the intensity of the emission lines.
Spectrograms of X~Per from 1913 to 2020 show X Per to have bright hydrogen
lines that are variable in structure, velocity, and intensity
\citep[e.g.][]{1972PASP...84..834C, 1998MNRAS.296..785T}.   % (e.g. Cowley et al. 1972, Telting et al. 1998). 
The primary is a rapidly rotating main sequence Be star
which forms an outwardly diffusing gaseous disc.
It has projected rotational velocity  $\approx 215$ km s$^{-1}$ 
\citep{1997MNRAS.286..549L}, 
and is classified as O9.5 \citep{1992A&A...259..522F},  
B0Ve \citep{1997A&A...322..139R},
B1Ve \citep{2019A&A...622A.173Z}.

In recent decades, X~Per exhibited two disc-loss episodes (around 1977 and around 1989)
when the emission lines and the circumstellar disc  were missing 
\citep[][]{1997A&A...322..139R,   % (Roche et al. 1997).
2001A&A...380..615C}.   % (Clark et al. 2001).  
The long-term variability of the emission lines and optical brightness 
indicate variable mass ejections from the donor star  into the circumstellar disc 
\citep{2014AJ....148..113L}.  %    (Li et al. 2014). 

We present optical spectroscopic observations obtained in the last two years  
and discuss asymmetries in the circumstellar disc and the appearance of eccentric wave.

%----------------------------------------------------------------------------- 
 \begin{figure*}     
  \vspace{12.0cm}   
  \includegraphics{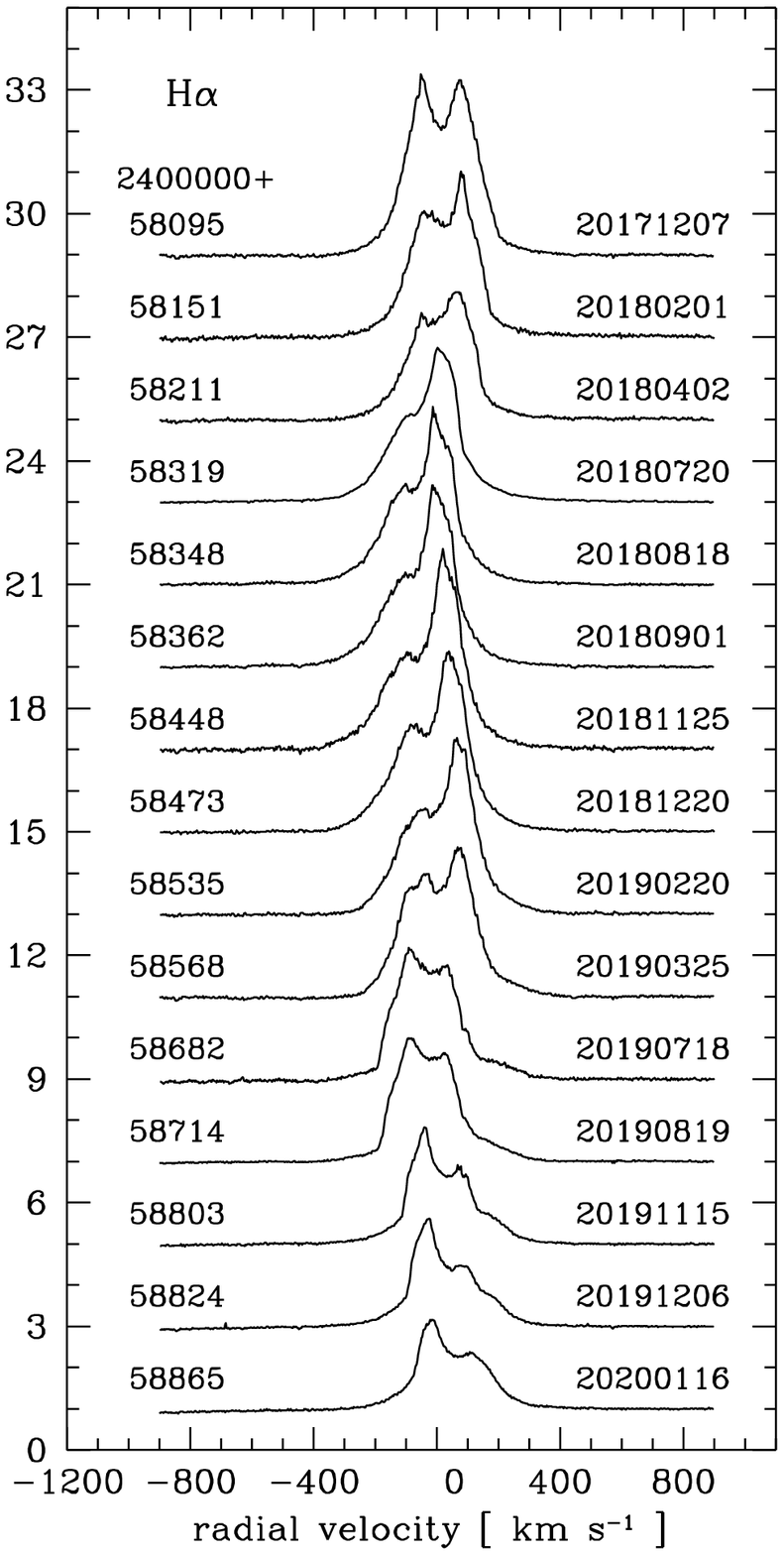}	  
  \includegraphics{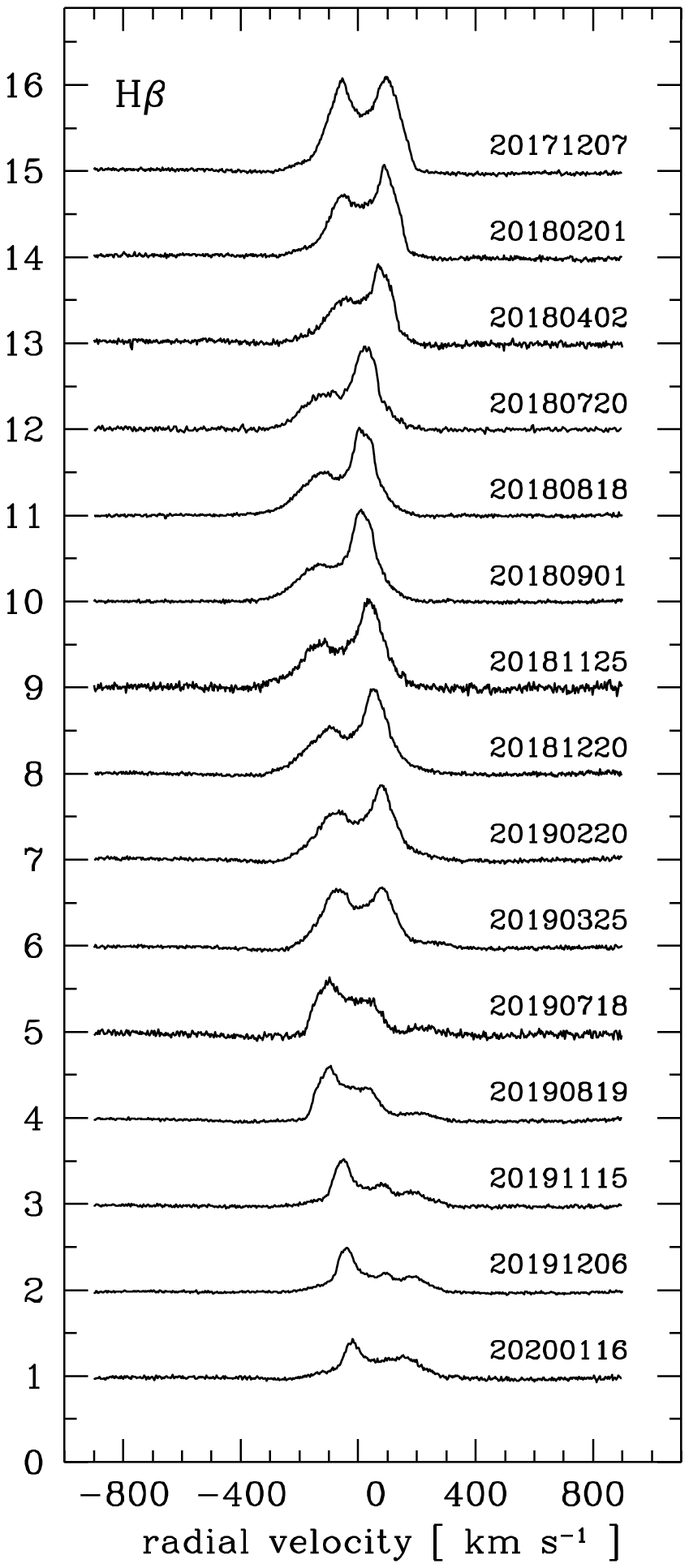}	  
  \includegraphics{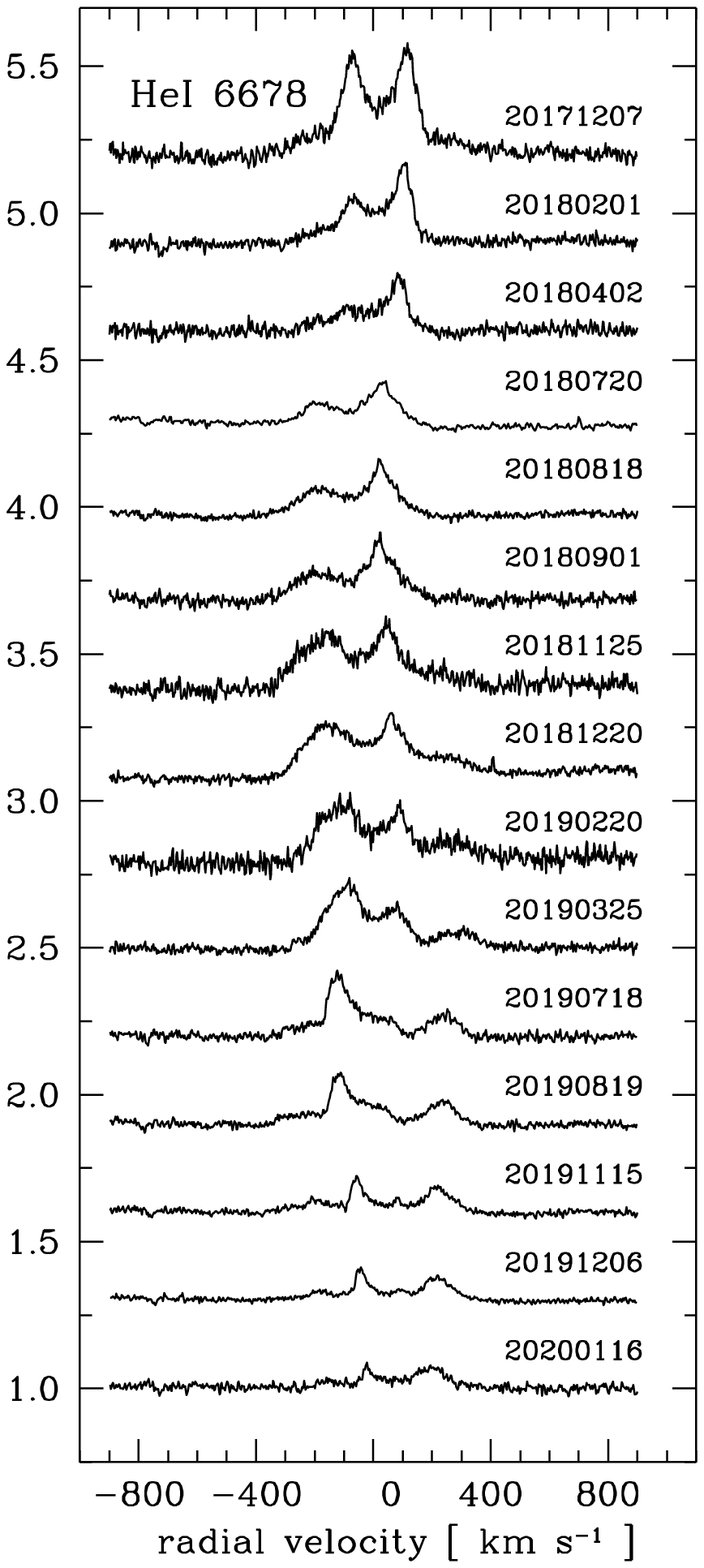}	  
  \caption[]{Variability of $H\alpha$, H$\beta$ and HeI~6678  emission line profiles of X Per.
	     The date of observations is in the format YYYYMMDD. }
\label{f1.pro}      
\end{figure*}	     
%------------------------------------------------------------------------------

\section{Observations}

We have 101 optical spectra of X Per on 57 nights secured with the 
ESpeRo Echelle spectrograph \citep{2017BlgAJ..26...67B}    %(Bonev et al. 2017)
on the 2.0 m RCC telescope of the Rozhen  National Astronomical Observatory, Bulgaria
%Usually,  we have one 2 minutes exposure (because $H\alpha$ could  be saturated on longer exposure)
%and one 10 or 15 minutes exposure, where  H$\beta$ and  HeI line can be investigated.
and  
with the HEROS spectrograph  \citep{2014AN....335..787S}   % (Schmitt et al. 2014)
on the 1.2~m  TIGRE telescope in the astronomical observatory La Luz in Mexico.
The variability of  $H\alpha$,  H$\beta$ and HeI~6678 emission lines
of X Per is presented in Fig.~\ref{f1.pro}. The spectra are normalized 
to the local continuum and a constant is added to each spectrum.
In this figure are plotted only 15 out of 101 spectra obtained. 
% In this figure is visible that 
% the blue peak gradually decreased in the period July - September 2018 
% and after that increased again. Third peak appeared in HeI and H$\beta$
% emission in 2019. 

On the spectra we measure the following parameters: equivalent width of the line, 
radial velocities of the peaks, intensity of the peaks. 
To measure the radial velocity we applied Gaussian fitting at the top of the peak. 
On a few spectra we see three peaks -- one violet and two red peaks.
In such cases we use the stronger red peak to calculate the distance between the peaks 
and V/R ratio. The measurements are given in a few tables in the Appendix. 

% How do you measure the peak distances? You should define which are the V and R peaks. 
% The third component should not be considered as a peak and should be treated separately.

\section{Results}
During the period December 2017  - January 2020, 
the equivalent width of  the $H\alpha$ emission line 
($W\alpha$) varies in the range  $-26$ \AA\ $ \le W\alpha \le -10$~\AA,
the equivalent width of the $H\beta$ emission line  
($W\beta$) varies in the range  $-4.4$~\AA\ $ \le W\beta \le  -1.4$~\AA,
and the equivalent width of the $HeI 6678$ emission line 
varies in the range  
$-2.0$~\AA\  $ \le W(HeI6678) \le -0.4$~\AA. 
The variability of the equivalent widths is presented in Fig.~\ref{f.EW},
together with V band magnitude and X-ray flux in  2 keV  -- 10 keV.
The V-band data are from  the American Association of Variable Star Observers (AAVSO). 
The X-ray data are from  MAXI  \citep{2009PASJ...61..999M}.
The vertical dashed lines indicate the three periods discussed in Sect.~\ref{V/Rratio}.  
There is a correlation between equivalent widths of these three lines.
The correlation analysis between 
$W\alpha$ and  $W\beta$, $W\alpha - W(HeI6678)$, $W\beta - W(HeI6678)$ gives 
correlation coefficient $\ge 0.80$, significance $p-value <  10^{-15}$,  
in other words a very strong correlation between the equivalent widths of the three lines.

It can be seen from Fig.~\ref{f.EW} that when the $W_\alpha$ decreases the V band brightness 
of X Per increases. The connection between  $W_\alpha$ and V band magnitude 
might be caused by the mass ejection from the Be star as discussed in  Sect.~4.1 by \cite{2014AJ....148..113L}.
There are no large changes in the X-ray flux. 
This indicates that the variability of the emission lines during the period
December 2017 - January 2020 does not affect the mass accretion rate onto the neutron star. 
  
%  Radii estimation through Huang+1972 method, is appropriate for symmetric profiles. 
%  You should mention that and say it is an approximation. 
%  You could also use the relation between W? and radius by Grudstrom & Gies 2006 and compare, 
%  although in this case, it looks like W? is not tracing the radius of the circumstellar disc. 
%  From Fig. 2 and 3 it looks like, at least before the appearance of the third component,
%   W? and ?V are correlated. How would you explain this? 

% 0.8 to 1.0 very strong relationship
% 0.6 to 0.8  strong relationship
% 0.4 to 0.6  moderate relationship
% 0.2 to 0.4  weak relationship
% 0 to 0.2    weak or no relationship

% 0 - 0.2 - very poor or very weak
% 0.2 - 0.4  - poor or weak
% 0.4 - 0.65 - fair or moderate
% 0.65 - 0.85 - strong or high
% 0.85 - 1.0  - very strong / high 
% 1.0  - perfect 

\subsection{Disc size}
\label{sec.Huang}

\begin{figure}   
 \vspace{9.7cm} 
  \includegraphics{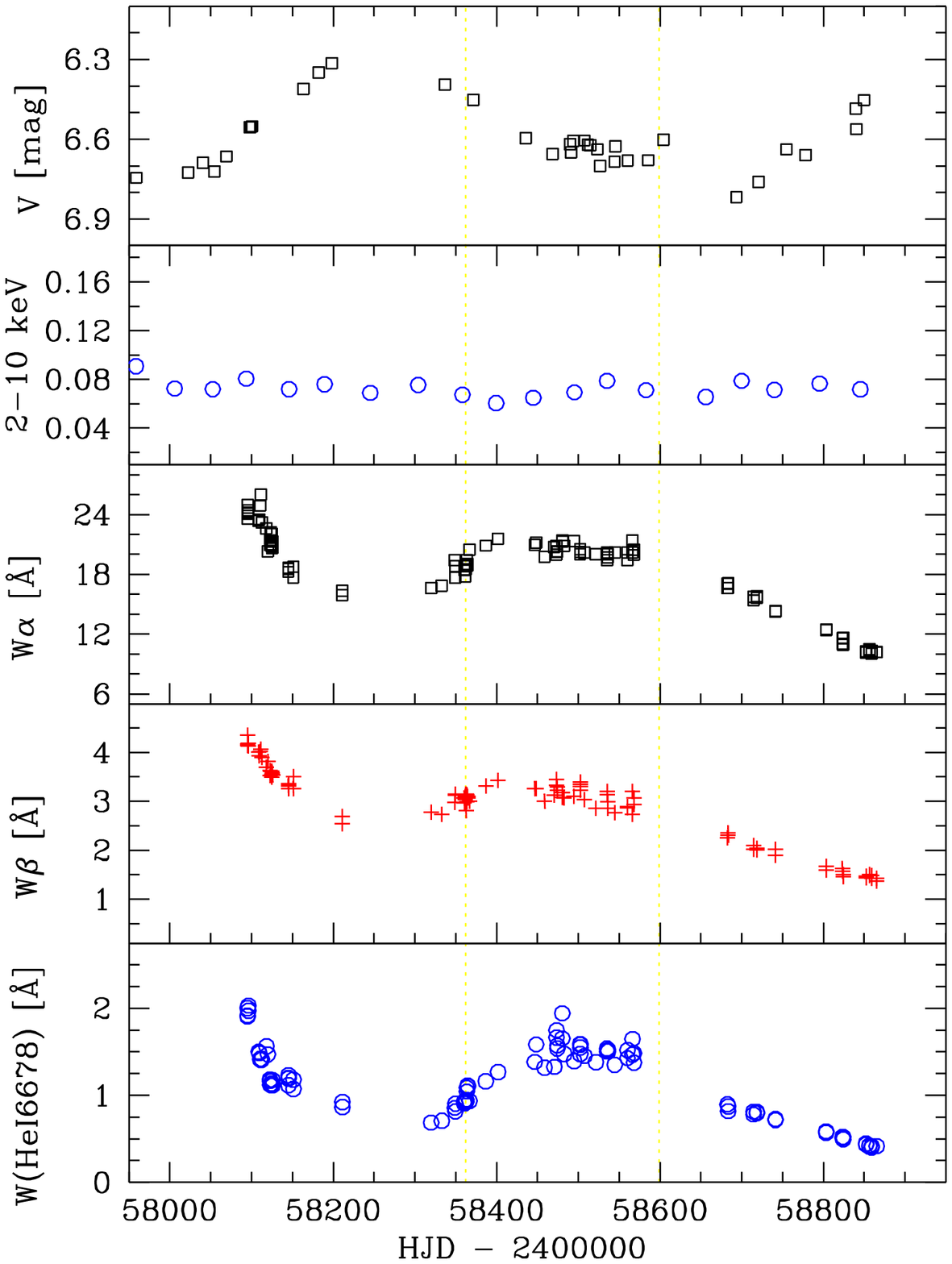} 	  
  \caption[]{Variability of X Per in V band (AAVSO),
     2-10 keV X-rays (MAXI), and the equivalent widths of  the emission lines $H\alpha$ (black squares), 
  $H\beta$ (red pluses), and HeI6678 (blue circles).  }
   \label{f.EW}
%-------------- 
  \vspace{6.5cm} 
  \includegraphics{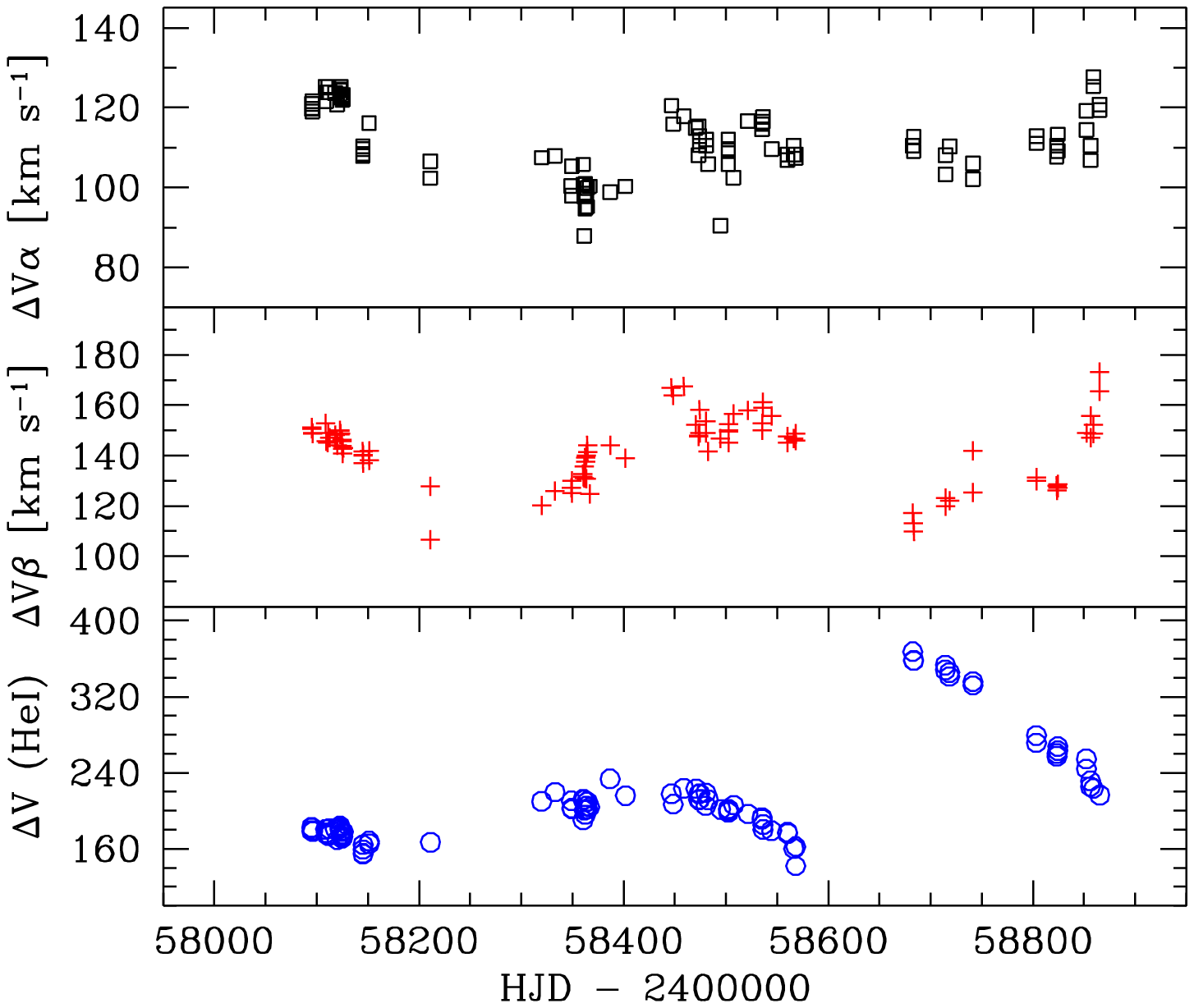}	  
  \caption[]{The distance between the peaks of the emission lines --
    $H\alpha$ (black squares), $H\beta$ (red pluses), and HeI6678 (blue circles).  }
   \label{f.dV}
%-------------- 
  \vspace{4.2cm} 
  \includegraphics{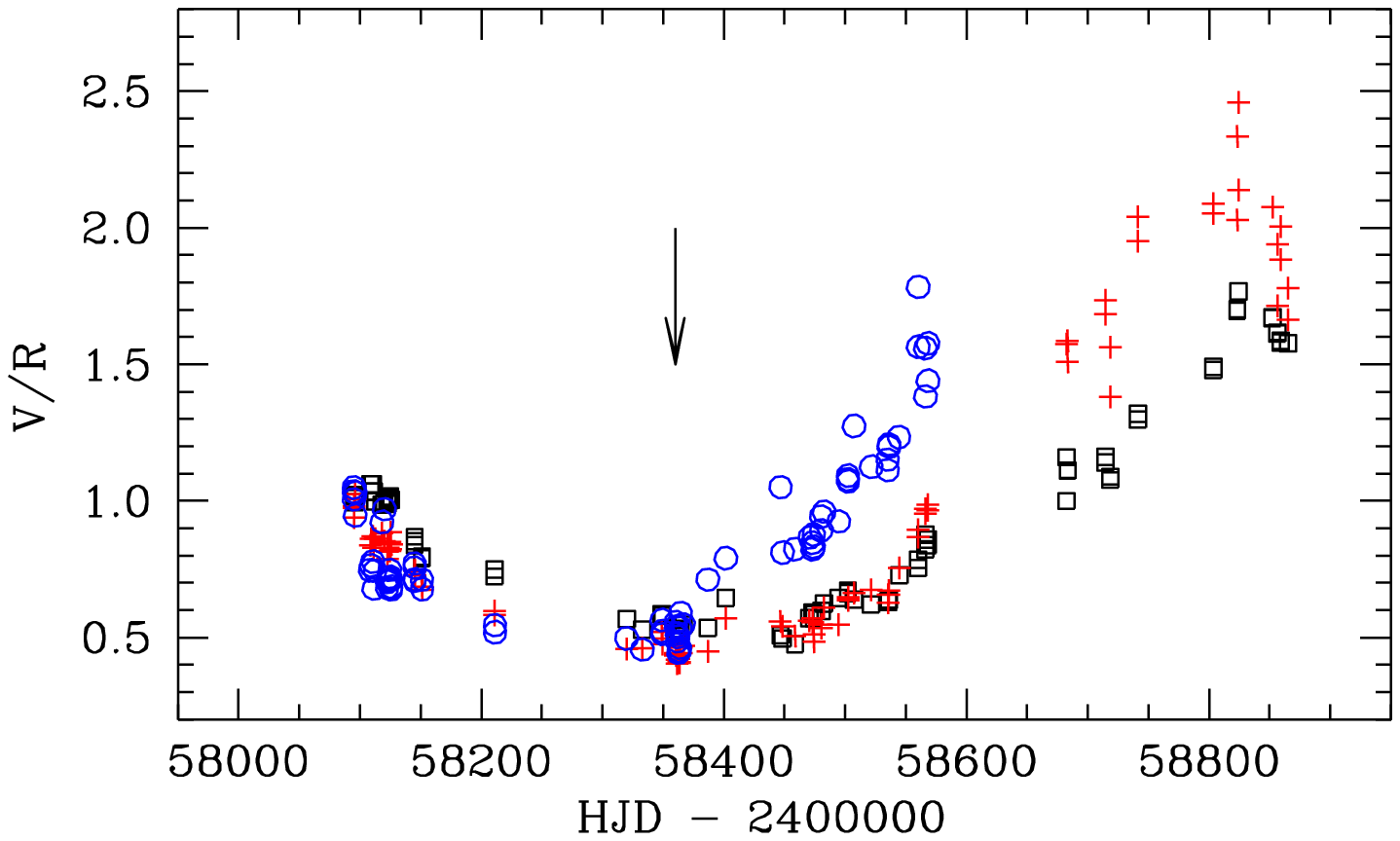} 	  
  \caption[]{V/R ratio for  $H\alpha$ (black squares), $H\beta$ (red pluses), and HeI6678 (blue circles).
   The arrow indicates the appearance of the eccentric wave.
   }  
\label{f.VRratio}      
\end{figure}

The variability of the distance between the peaks is presented in Fig.~\ref{f.dV}. 
The emission lines form  in the disc surrounding the Be star.  
% and the total flux of the feature (measured as the line equivalent width) is closely related to the size of the disc. 
The discs of the Be stars are Keplerian supported by the rotation 
[e.g. \cite{2013A&ARv..21...69R}, \cite{2016ASPC..506....3O} and references therein].     % 2013A&ARv..21...69R,  Rivinius Carciofi Martayan[e.g. 
For a Keplerian circumstellar disc the peak separation can be regarded as a measure of 
the outer radius ($R_{disc}$) of the emitting disc \citep{1972ApJ...171..549H}:   % (Huang 1972):  
%-----------------------------------------------------------
 \begin{equation}
  R_{disc} = R_1 \frac{ (2\,v\,\sin{i})^2} {\Delta V ^2 }, 
  \label{Huang}
  \end{equation}
%------------------------------------------------------------
where  $R_1$ is the radius of the primary and $v\,\sin{i}$ is its projected rotational velocity. 
% {\bf  \textcolor{red}  
Radii estimation through this method, is a good approximation for symmetric profiles. 
The projected rotational velocity of the primary 
is estimated  to be  $v \: \sin \: i =200$~km~s$^{-1}$  \citep{1982ApJS...50...55S},   % (Slettebak 1982),
$v \: \sin \: i =215 \pm 10$~km~s$^{-1}$  using the HeI $\lambda 4026$ \AA\  absorption line 
\citep{1997MNRAS.286..549L}  % (Lyubimkov et al. 1997)
and  $ v \: \sin \: i = 191 \pm 12$~km~s$^{-1}$
from the width of the $H\alpha$ emission  \citep{2019A&A...622A.173Z}.     %  (Zamanov et al. 2019).
For the primary we adopt  $R_1=10.5$~$R_\odot$ \ and mass  $M_1=13.5 \: M_\odot$
\citep{2019A&A...622A.173Z}.   %  (Zamanov et al. 2019).
This gives the Keplerian velocity on the surface of the star $V_{Kepl} = 495$~km~s$^{-1}$. 
Adopting inclination $i=30^0$ \citep{2001ApJ...546..455D},  % (Delgado-Mart{\'\i} et al. 2014), 
and that the star rotates with 0.8 of the critical velocity
\citep[e.g.][]{2003PASP..115.1153P},  % (e.g. Porter \& Rivinius 2003),
we estimate $v \: \sin \: i  \approx 198$~km~s$^{-1}$, which is in agreement with the above values. 
This agreement also is a clue that there is no considerable deviation between the orbital plane of the binary
and the equatorial plane of the Be star.
If it exists at all it should be less then $5^\circ$.  

On the spectrum  20190325  (see Fig.~\ref{f1.pro}),  
the HeI emission extends upto $\approx 370$~km~s$^{-1}$  at zero intensity.  
We note that velocities in the profiles of the emission lines 
$> 495 \sin i$~km~s$^{-1}$, which is $\approx 250$~km~s$^{-1}$,   
are probably super-Keplerian velocities, and indicate  eccentric motion in the disc, 
where these parts of the emission line are formed (see Sect.~\ref{sec.HeI}).

In December 2017 the emission lines have double-peaked symmetric profiles
(see Fig.~\ref{f1.pro}). 
For 5 spectra obtained in December 2017 we measure
$\Delta V\alpha =120 \pm 1  $~km~s$^{-1}$,
$\Delta V\beta  =150 \pm 1  $~km~s$^{-1}$, and
$\Delta V_{HeI} =180 \pm 2  $~km~s$^{-1}$.
Using Eq.~\ref{Huang}, these values correspond  to disc size  
$R_{disc} (H\alpha) = 134 $~$R_\odot$, 
$R_{disc} (H\beta) =  86  $~$R_\odot$, and 
$R_{disc} (HeI6678) = 60  $~$R_\odot$. 
During the period of our observations, 
the average distance between the peaks 
of the lines is 
$\Delta V\alpha = 111 \pm 14 $~km~s$^{-1}$,
$\Delta V\beta  = 142 \pm 13 $~km~s$^{-1}$,
$\Delta V_{HeI} = 213 \pm 52 $~km~s$^{-1}$.
%111	158
%142	96
%210	44
%220	40
%the two peaks are visible in the emission lines profiles
%and we can estimate the disc radius using Eq.~\ref{Huang}.
These values correspond to the
following average disc size for different emission lines (calculated using Eq.~\ref{Huang}): 
$R_{disc} (H\alpha) = 156 $~$R_\odot$, 
$R_{disc} (H\beta) =  96  $~$R_\odot$, 
$R_{disc} (HeI6678) = 43  $~$R_\odot$. 
The typical errors are about $\pm$ 5 \%.
For the calculations of the outflowing velocity (Sect.~\ref{sect.Vout})
we will use $R_{disc} (H\beta) = 86 - 96$~$R_\odot$. 
The average ratios between disc sizes (which is equivalent to the ratio of the peak separations)  are: 
$ R_{disc} (H\alpha) / R_{disc} (H\beta)    = 1.62  $,
$ R_{disc} (H\alpha) / R_{disc} (HeI6678)   = 3.63  $.

%
%  HeI      6678.1517
%  Halpha   6562.817
%  Hbeta    4861.298
%

%----------------------------------------------------------------------------- 
%------------------------------------------------------------------------------

%----------------------------------------------------------------------------- 
 \begin{figure}   
  \vspace{6.4cm} 
  \includegraphics{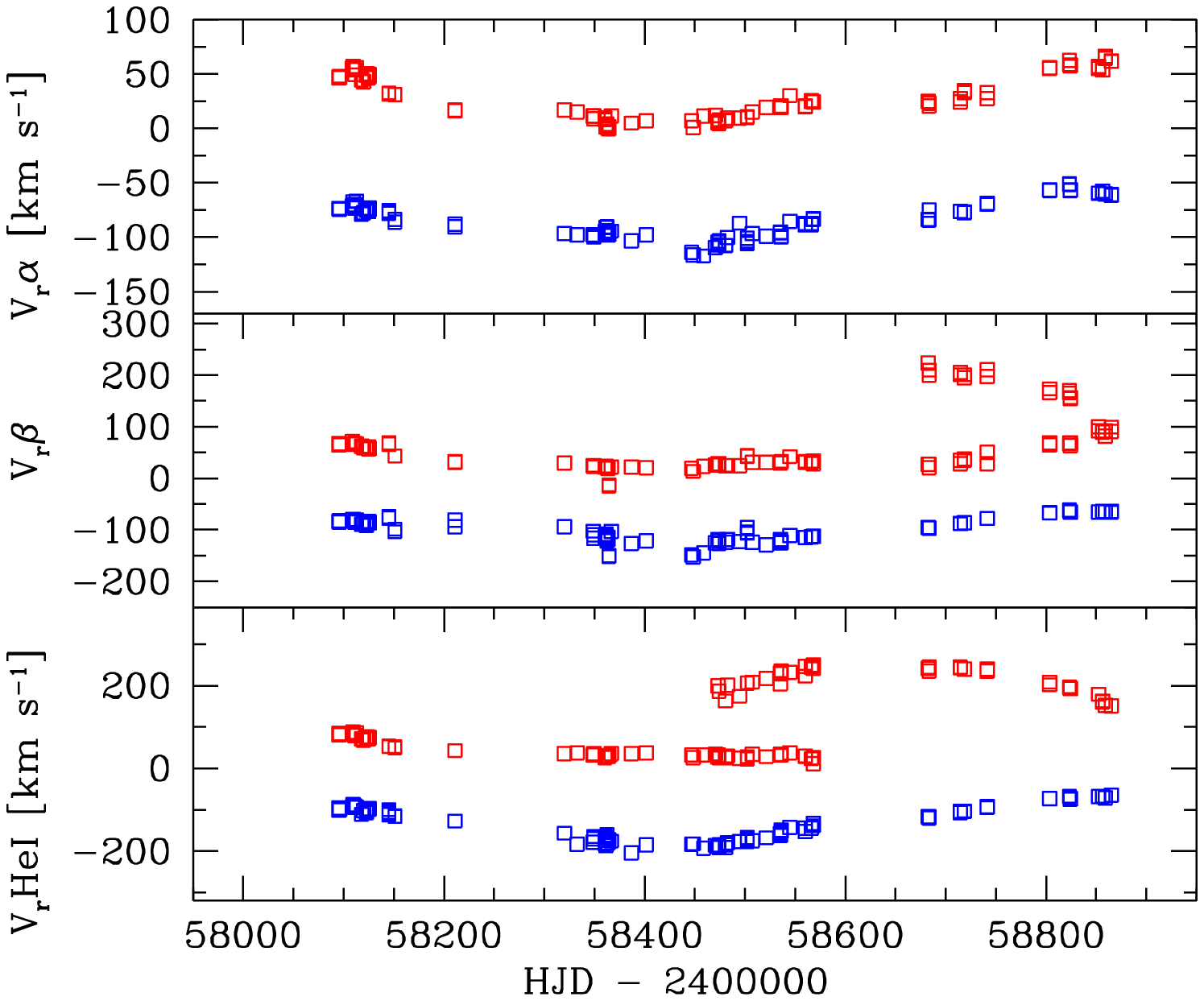}	  
  \caption[]{ Radial velocities  of the peaks.   }
  \label{f.Vr}

  \vspace{7.4cm} 
  \includegraphics{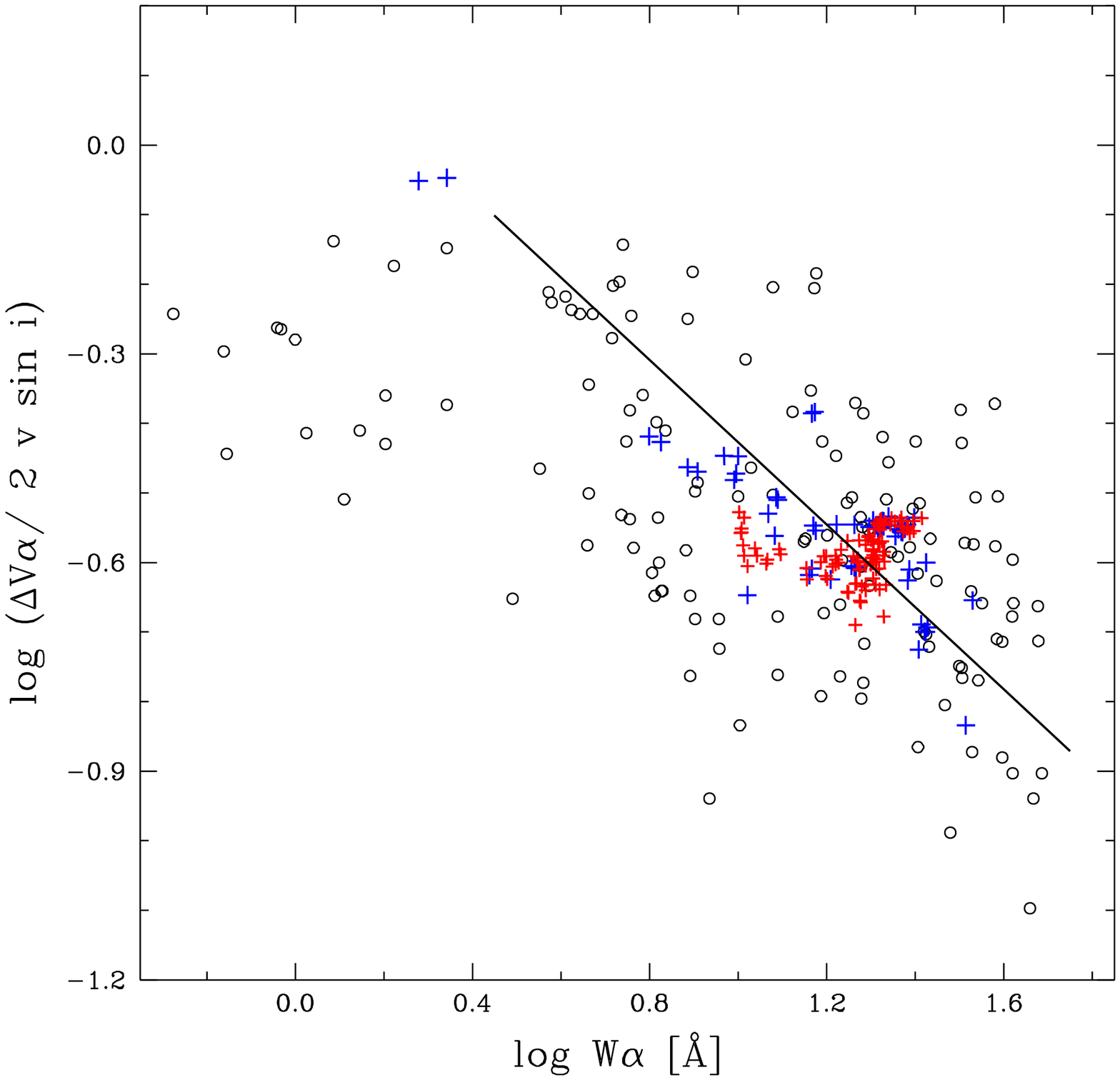}	  
  \caption[]{ $\Delta V\alpha$ versus $W_\alpha$. The circles are the Be stars (see text)
  and the red crosses are our observations of X~Per. }
  \label{f.Wa.dV}
 
  \vspace{6.9cm} 
  \includegraphics{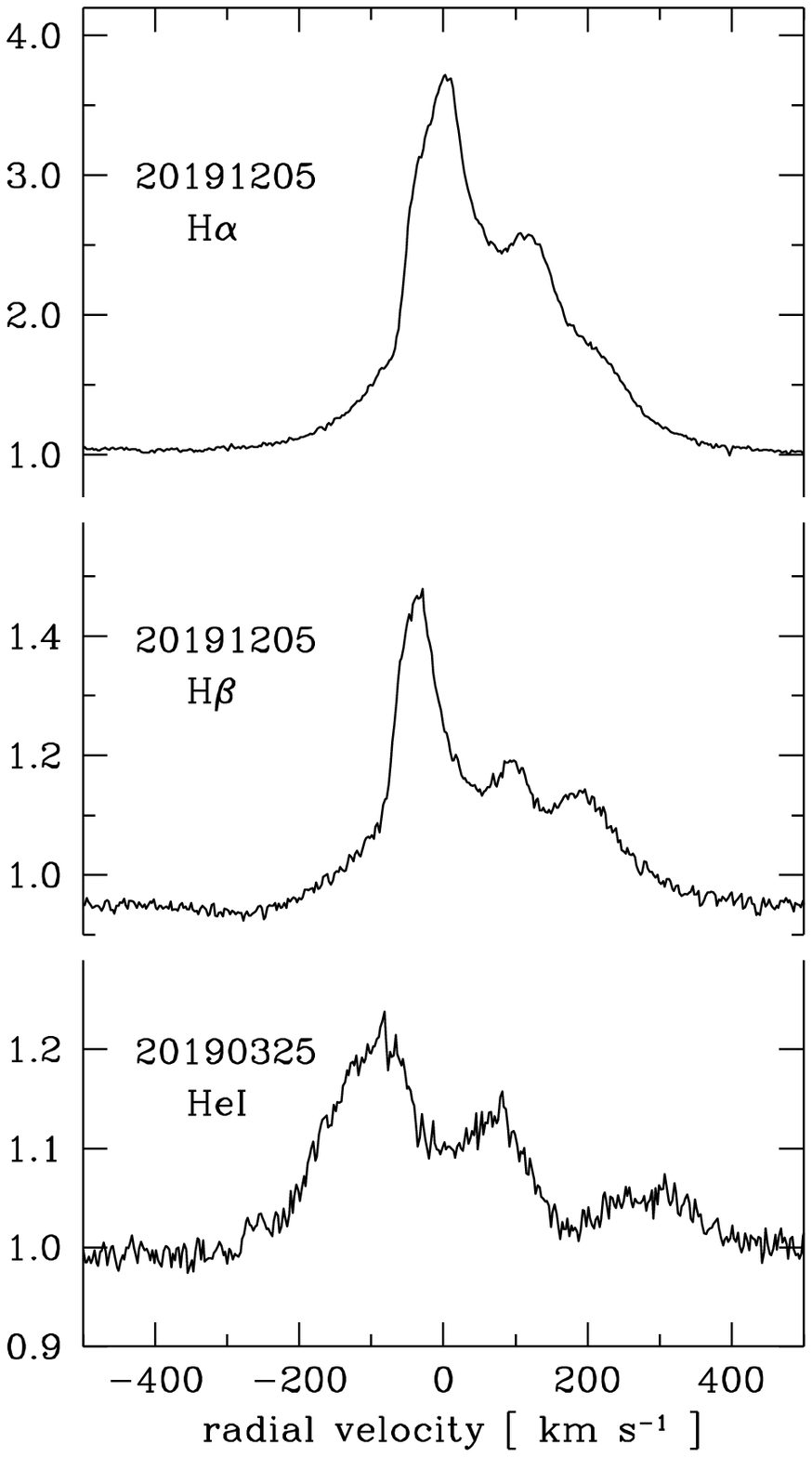}	  
  \includegraphics{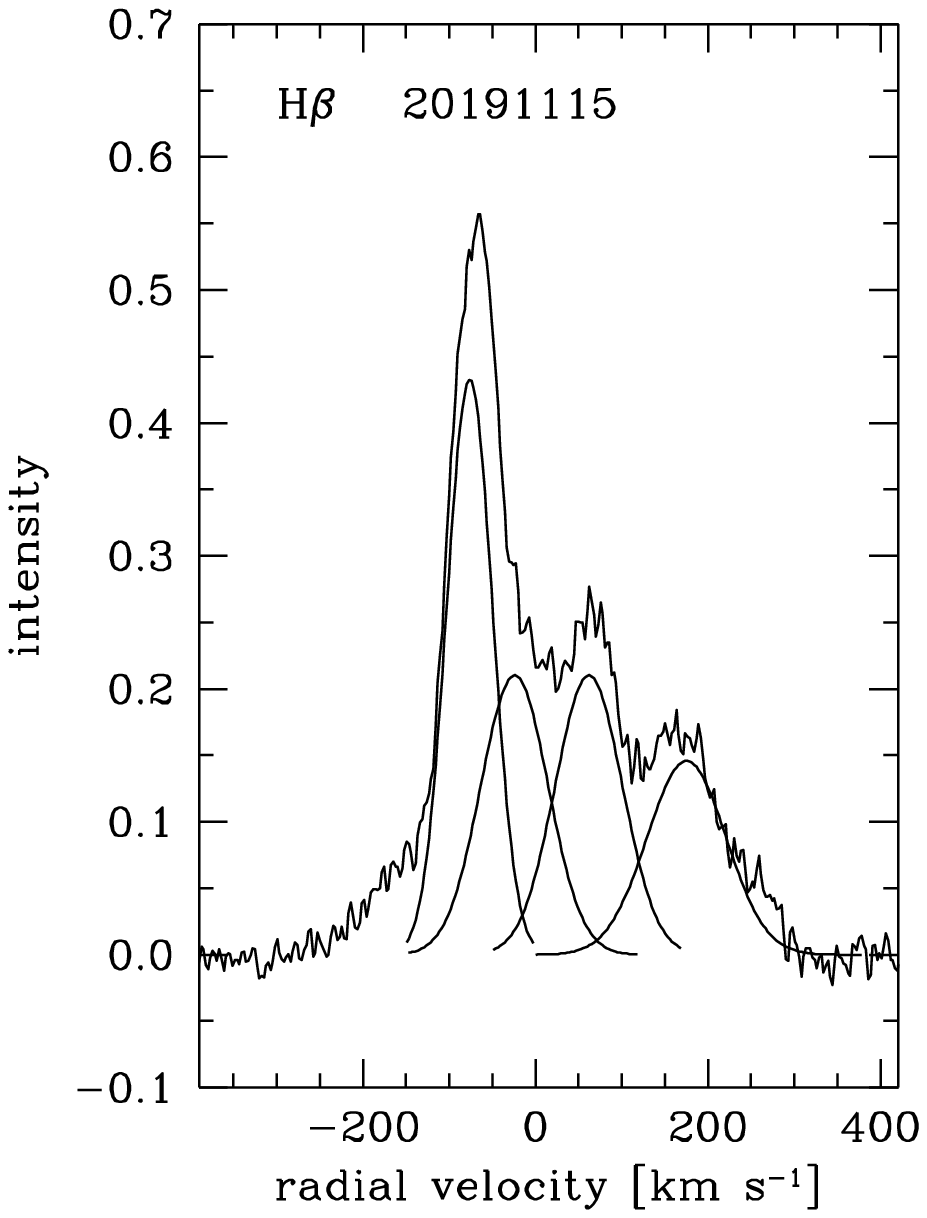}	  
  \caption[]{The left panel shows the three component structure of HeI~6678 and $H\beta$ lines, and
  red shoulder in the $H\alpha$  line. Right panel is an approximation with four Gaussian (see Sect.\ref{e.wave}). 
  }
  \label{f.3comp}
\end{figure}  
%-------------------------------------------------------------------------------

\subsection{V/R ratio and  asymmetries in the disc}
\label{V/Rratio}

In Fig.\ref{f1.pro} it is seen that the  emission lines are 
symmetric  in December 2017, with  $V/R$ ratio $\approx 1$. 
After that  all three lines become asymmetric.
Fig.~\ref{f.VRratio} presents the V/R ratio, calculated as $V/R=(I_V-1)/(I_R-1)$, where $I_V$ and
$I_R$ are the intensity of the blue (violet) and red peak, respectively. The spectra are normalized before 
the measurements and the continuum level is $\equiv 1.0$.

(1)   During the first part of the period (MJD~58095 - MJD~58365) 
in all three lines the V/R-ratio varies in practically the same manner --
it decreased from $V/R  \approx 1$ to  $V/R  \approx 0.5$, 
with  speed  $\frac {\Delta (V/R)}{\Delta t} \approx 2 \times 10^{-3}$ d$^{-1}$.
The fact that the changes in V/R-ratio are practically identical in 
all three lines ($H\alpha$, $H\beta$ and HeI,
see Fig.\ref{f.VRratio})
means that the asymmetry develops in the entire disc simultaneously.

(2) During the second  part  of the period 
(MJD58365 - MJD58600) the V/R-ratio of $H\alpha$ and 
$H\beta$ goes up from 0.5 to $\sim 1.0$. 
However the behaviour of the V/R-ratio of HeI6678  deviates from the behaviour of the H$\alpha$
and $H\beta$ lines (see Fig.~\ref{f.VRratio}).  V/R (HeI) changes from  0.5 to 1.5, having a faster rate  
of change $\frac {\Delta (V/R)}{\Delta t} \approx 4 \times 10^{-3}$ d$^{-1}$. 
For comparison, the  H$\alpha$ emission peaks 
have rate of change $\frac {\Delta (V/R)}{\Delta t} \approx 1.7 \times 10^{-3}$ d$^{-1}$
and the $H\beta$ peaks -- $\frac {\Delta (V/R)}{\Delta t} \approx 2.3 \times 10^{-3}$ d$^{-1}$ .
In addition to that, a third component appeared in the red side of the HeI profile
(Fig.~\ref{f.3comp}). This third component emerged in HeI in November 2018. 
% On spectrum obtained  on 25 March 2019
% $V/R(H\alpha) < 1$, $V/R(H\beta) \approx 1$,  $V/R(HeI6678) > 1$
% (see Fig.\ref{f1.pro}). 
% HeI6678 has a peculiar three peak profile (Fig.\ref{f.3comp}).
This  is an indication 
that  the structural changes begin in the innermost region of the disc. 

(3) The behaviour of the V/R-ratio of $H\beta$ is similar to that of $H\alpha$
till MJD~58600. After  it it begins to deviate (see Fig.~\ref{f.VRratio}). 
This is an indication that the asymmetry spreads out in the disc.

The development of asymmetries visible in the $H\alpha$ peaks corresponds to processes in the outer parts,
while the variability of HeI is connected with changes in the inner parts of the circumstellar disc.

\subsection{Third component in He~I}
\label{sec.HeI}

In November 2018 a third component appears in the red side of the HeI line.
In March 2019, this component peaks at  radial velocity  $\sim 250$~~km~s$^{-1}$, 
and at zero intensity  it is in the range 130-390~km~s$^{-1}$. 
The three component structure is demonstrated on Fig.~\ref{f.3comp},  left panel.
The Keplerian velocity around the primary is
$V_{Kepl} = \sqrt{ G M_1 /r }$,
 where $G$ is the gravitational constant and $r$ is the distance. 
%Using these values, 
We estimate that the observed velocity of the third component  corresponds to
a distance from the centrum of the primary of 40 - 20 $R_\odot$, respectively. 
Bearing in mind that $R_1 \approx 10.5$~$R_\odot$, 
it means that this component 
originates somewhere  about 1-3 stellar radii above the stellar surface. 
The velocity of this component is similar to the velocities
observed during the double disc formation in 1994 
\citep{1995MNRAS.276L..19T},     %(Tarasov \& Roche 1995),
when the blue and red peaks of the inner disc were
at $-308$ km~s$^{-1}$ and $+258$~km~s$^{-1}$, respectively.  

%  Tarasov \& Roche (1995) reported the interesting formation of a double circumstellar
%  disc in X Per, inferred from the quadruple emission peak structure in HeI~6678 line.
%  eccentric waves in astrophysical discs  Lynch \& Ogilvie

\section{eccentric wave}

The discs around stars, or other central massive bodies,
can support disturbances in which the fluid motion is nearly Keplerian with non-zero eccentricity.
The eccentric waves are expected not only in Be discs but also 
in accretion discs, protoplanetary discs, and in
discs with an embedded planet (e.g. \cite{2019MNRAS.488.1127L} and references therein).   % Lynch \& Ogilvie 2019
In X~Per, the appearance of a significant density enhancement in the disc near the stellar surface 
and its evolution was also observed and discussed  by \cite{1995MNRAS.276L..19T}
and 
\cite{2001A&A...380..615C}. 
%The behaviour of the emission lines 
%inner eccentric density wave similar to the observed by T&R1995 and Clark+2001
Assuming  that the observed behaviour of the emission lines of X~Per is due to 
an eccentric wave, we calculate its velocity and eccentricity.  

\subsection{Velocity of the eccentric wave}
\label{sect.Vout}
We first see the appearance of the third component in HeI lines on our spectrum in November 2018  (20181124).
It is not visible on our spectrum obtained in mid-October 2018 (20181010). 
In March 2019 this component starts to be visible in the red side of $H \beta$. 
The profile of $H \beta$ on 5 December 2019 (20191205) with three peaks 
is  similar to that of HeI obtained 8 months earlier (20190325), see Fig.\ref{f.3comp}. 
We suppose that the eccentric wave began somewhere about MJD~58360 (about end of August 2018)
when the behaviour of the HeI peaks started to deviate from H$\beta$ and H$\alpha$ 
(see the arrow on Fig.~\ref{f.VRratio}). 
At this time it is likely to be at a distance $r \ge R_1$.  
It reached  near %to the outer edge of the HeI disc probably in June  2019 (when the two red peaks of HeI6678 blended), 
to the outer edge  of the $H\beta$ disc in January 2020
(when the two red peaks of $H\beta$  blended).
This indicates that the eccentric wave spreads from inside out.
Bearing in mind the disc size in the lines
(Sect.~\ref{sec.Huang})
this corresponds to an  outflowing velocity of the wave 
$v_{wave} \approx 1.2 \pm 0.1$~km~s$^{-1}$.

%  \lesssim      \gtrsim

In the above calculation we supposed that  the motion of the density wave is linear. 
We also supposed that the peak merging corresponds 
to the moment when the wave reached the outer edge.
We performed numerical experiments with Gaussian fitting of 
the peaks and their blending (merging). 
The experiments showed that to observe two peaks the difference between 
their velocities must be $\gtrsim  30$~km~s$^{-1}$. If it is less than 30~km~s$^{-1}$ then we will see one peak. 
This value depends  on the signal-to-noise ratio of the spectra, width of the peaks, their intensity, etc.
This limit of  30~km~s$^{-1}$ was found with numerical experiment with two Gaussian peaks, 
having width and intensity similar to those observed. 
This also indicates that the above estimation should be considered as an upper limit 
of  $v_{wave}$.
Assuming that the merging of the peaks corresponds to the moment when 
 the wave is at $0.85 \pm 0.15 $ of the disc size, this will 
give $v_{wave} \approx 1.1 \pm 0.2$~km~s$^{-1}$.

In Fig.~\ref{f.Vr} are plotted the heliocentric radial velocities of the peaks
for $H\alpha$  ($V_r \alpha$),  for $H\beta$ ($V_r \beta$), and for HeI~6678 ($V_r HeI$). 
The radial velocities of the HeI peaks indicate that the wave begins somewhere about MJD~58350 (August 2018). 
The radial velocities of the $H\beta$ peaks  indicate that the two red peaks  
merged somewhere about MJD~58850 (January 2020). 

% this can correspond to a the moment when the
% wave is at about  50\% ?  75\%
% corresponds to the moment when the wave is of about  peaks 2 and 3 get blended
% If we suppose that 
% The peak merging (blending): 
% in H-beta - to observe two peaks the difference between 
% their velocities must be $> 50 km/s$. If it is less than 50~km~s$^{-1}$ then we will see one peak
% It depends also on the signal-to-noice ratio of the spectra, width of the peaks, their intensity, etc.
% This limit  50~km~s$^{-1}$ is found with numerical experiment with two gaussian peaks, 
% having width and intensity similar to the observed. 
% In the spectrum of X~Per four peaks were observed in HeI6678 emission line in 1994-1995.
% This is interpreted as due to a double disc structure (Kunjaya \& Hirata 1995; Tarasov \& Roche 1995). 
% Probably, we have similar situation in March 2019, however the blue peak of the inner disc
% coincides with the blue peak of the outer disc (on Fig.\ref{f3.HeI} is visible that 
% the blue peak is wider and stronger). 
% If this is the case, 
% the inner disc is likely to be elliptical. 
% For the instantaneous orbital speed of a body at any given point in its trajectory

In Fig.~\ref{f.Wa.dV} we plot distance between the peaks of $H\alpha$ normalized with the stellar 
rotation  versus $W\alpha$. These two parameters correlate in the Be stars, representing
the fact that the disc grows as $W_\alpha$ increases. 
The black open circles are data for Be stars taken from 
\citet{1983A&AS...53..319A},    %   Andrillat (1983),             OK
\citet{1986A&A...166..185H},    %   Hanuschik (1986),             OK
\citet{1988A&A...189..147H},    %   Hanuschik et al. (1988),      OK
\citet{1992A&AS...95..437D},     %   Dachs et al. (1992),         OK
\citet{1992ApJS...81..335S},      %   Slettebak et al. (1992)     OK
and \citet{2013A&A...550A..79C}.   %  and Catanzaro (2013).       OK
The red plus signs are our measurements of X~Per. 
We see that X~Per is close
to the average behaviour of the Be stars. This indicates that the eccentric wave does not change the 
overall structure of the circumstellar disc.   
%  In this figure are plotted  138 data points for Be stars 

%---------------------------------------------
\begin{table}
  \centering
  \caption{  %Radial velocities of the peaks on a few spectra. 
   In the table are given date of observations
  (in the format YYYYMMDD), heliocentric radial velocities of the three peaks, 
  and the calculated eccentricity of the eccentric wave.  }
   \begin{tabular}{lclrcccccr} 
   \hline
   \\
 date-obs & $V_{r1}$ &  $V_{r2}$ &  $V_{r3}$ & $e_{wave}$ & \\
 \\
 HeI 6678 \\
 20190221 &  -147.2 &  29.9 &  231.9 &  0.17  & \\
 20190323 &  -137.7 &  24.8 &  247.8 &  0.23  & \\
 20190325 &  -133.4 &  23.9 &  246.3 &  0.24  & \\
 20190220 &  -155.1 &  29.8 &  249.0 &  0.18  & \\
 \\
 $H\beta$ \\
 20190819 &  -89.6  &  33.0 &  210.2 &  0.33  & \\
 20190822 &  -86.6  &  39.2 &  201.5 &  0.32  & \\
 20191115 &  -67.0  &  62.0 &  176.2 &  0.36  & \\
 20191206 &  -65.8  &  63.6 &  157.6 &  0.31  & \\
 \\
 $H\alpha$  \\
 20191115 &   -57.1 &  59.8 &  145.2 &  0.33  & \\
 20191115 &   -56.9 &  55.0 &  148.6 &  0.36  & \\ 
 20191206 &   -63.5 &  52.3 &  155.6 &  0.32  & \\ 
\\
 \hline 						   
 \end{tabular}  						    
 \label{tab.3peaks}				  
 \end{table}					  
%------------------------------------------

\subsection{Eccentricity  of the eccentric wave}
\label{e.wave}

The vis-viva equation, connects 
the instantaneous orbital speed of a body at any given point in its trajectory
with distance. For periastron and apastron we have:
\begin{equation}
v_{per}  = \sqrt{ GM_1 \left( \frac{2}{a(1-e)} - \frac{1}{a} \right)} 
\label{eq.per} 
\end{equation}
%-----------
\begin{equation}
v_{ap}  = \sqrt{ GM_1 \left( \frac{2}{a(1+e)} - \frac{1}{a} \right) } 
\label{eq.ap} 
\end{equation}
where $a$ is the length of the semi-major axis of the elliptical orbit, 
and $a(1-e)$ and $a(1+e)$ are the distances at which the speed is to be calculated
(in our case periastron and apastron, respectively).

Using three spectra on which $H\alpha$ is symmetric 
(20161211, 20170317, 20171207) we measure the velocity 
at the half maximum of the $H\alpha$  emission  $-11 \pm 3$~km~s$^{-1}$. 
The measurement is done using the position of the bisector at the half maximum intensity 
of the emission, as shown in Fig.~1 of \cite{1986A&A...158..392G}. 
 The old spectroscopic data of Hutchings (1977) give an  average 
 value for the radial velocity of 
 the Balmer absorption $\approx -20$~km~s$^{-1}$ and for  HeI lines $-19 \pm  10$~km~s$^{-1}$.
\cite{2007ApJ...660.1398G}  %  Grundstrom et al. (2007) 
give systemic velocity 
$\gamma = 1.0 \pm  0.9$~km~s$^{-1}$ on the basis of
absorption lines in the IUE spectra.
% The value of Grundstrom et al. (2007)2007ApJ...660.1398G
% is based on many more measurements
% and should be more accurate. 
On the spectra where 3 peaks are visible we measure the heliocentric radial 
velocities of the peaks. They are given in Table~\ref{tab.3peaks}. 
For HeI and H$\beta$ these are  measurements of the peaks. 
For $H\alpha$ it is the velocity of the shoulder.  
Correcting these values for the systemic velocity: \\  % We assume  that  
$V_{r1} - \gamma  = v_{ap} \sin i$   \\
  and \\
$V_{r3} - \gamma = v_{per} \sin i$,    \\
% that bluest peak is connected with 
% Correcting these values for the systemic velocity $\gamma = 1.0$~km~s$^{-1}$,
and using Eq.~\ref{eq.per} and  Eq.~\ref{eq.ap}, we estimate 
the eccentricity of the wave, which is given in the last column of Table~\ref{tab.3peaks}. 
The average eccentricity of the wave is $e_{wave} = 0.29 \pm 0.07$. 

In the above calculations we adopted  $\gamma = -11$~km~s$^{-1}$.
If we assume   $\gamma = -1$~km~s$^{-1}$, the estimated $e_{wave}$ will increase 
up to  $e_{wave} =  0.36 \pm 0.07$. 
There should be two blue peaks corresponding to the two red peaks. In our spectra we see
only one, see Fig.~\ref{f.3comp} (left panel). It means that the two blue peaks are blended
(and have approximately equal velocities). 
To estimate the possible influence of the blending 
on our value of $e_{wave}$, we fitted the 
profile with four Gaussian (Fig.~\ref{f.3comp}, right panel). During the fitting 
we explicitly assumed that the two inner peaks are identical. 
The fitting suggests that the peak (corresponding to the wave) could be blue 
shifted with 10 -- 20~km~s$^{-1}$. A correction  of $-10$~km~s$^{-1}$ 
to the measured radial velocity  produces a lower value $e_{wave} = 0.24 \pm 0.06$.  
% The fitting indicates that the radial velocity we used is likely 
% should be corrected with 
These two sources of  uncertainty  give  the range $0.16 < e_{wave} < 0.42$
for the eccentricity of the wave.

\section{Discussion}

Eccentric discs, in which planet, or star, or fluid elements
follow elliptical orbits of variable eccentricity around a central mass,
have  applications in various astrophysical objects: Be stars \citep{2008MNRAS.388.1372O},
galactic nuclei \citep{2018MNRAS.480.2929C}, 
planetary rings and protoplanetary discs \citep{2019ApJ...882L..11L, 2020arXiv200605529M}.
A more deep understanding of the appearance and  the evolution of asymmetries and  eccentricities 
in the circumstellar disc of X~Per and other Be/X-ray binaries 
would therefore be of general interest.
There are different theories about the origin of the long-term  variability
observed in Be stars
[e.g. \cite{1995A&A...300..163H}, Section 4.2].   % [e.g. Hanuschik et al. (1995), Section 4.2]. 
The most accepted is the global oscillation scenario, 
which proposes that a Keplerian disc around a Be stars 
is subject to global distortion -- a one-armed density wave
\citep[][]{1991PASJ...43...75O,  1993A&A...276..409S, 2006A&A...456.1097P},  
which is an updated  and  sophisticated
version of the old elliptical disc model    \citep{1931ApJ....73...94S}.  %    (Struve 1931).
Eccentric mode in Be/X-ray binaries can be excited in the disc through direct driving 
as a result of a one-armed bar potential of the binary \citep{2002MNRAS.337..967O}.

In our opinion, during the first part or our observations (MJD~58095 -- MJD~58365) 
we observe
a global distortion ($H\alpha$, $H\beta$, $HeI$ vary together)
similar to that observed in  $H~1145-619$ (see Fig.~7 by
\cite{2017A&A...607A..52A}).   %   Alfonso-Garz\'on et al. 2017).
% global one-armed density wave which is visible in all three lines.   One-Armed Density Perturbation
During the period MJD~58365 --  MJD~58865, 
we observe an eccentric wave, which
starts in the innermost parts of the disc and spreads out with velocity 
$v_{wave} \approx 1.1$~km~s$^{-1}$. 
Once the wave is in the disc, it should begin to rotate. Its rotation 
is expected to  be much slower than the disc rotation \citep{1986Ap&SS.119..109O}.   
The radial velocities on Fig.~\ref{f.Vr} indicate that the period of rotation of the density wave 
is  probably $\sim 800$~d.

The interesting behavior of the HeI emission of X~Per was studied 
by \cite{1995PASJ...47..589K}
and 
by \cite{1995MNRAS.276L..19T}. 
They observed four peaks in HeI emission in 1995 and 
suggested the formation of a double circumstellar disc. 
\cite{1977PASJ...29..477H}
were first to argue about the existence of a two component structure of 
the circumstellar envelope around a Be stars following the example of the well known star Pleione (BU Tau). 
They proposed a model consisting of two layers 
and also considered a separate fast rotating layer closer to the stellar equator. 
The formation of a new envelope, coexisting with the previous one
is also detected by \cite{2010A&A...516A..80N} on the basis of four peak structure of 
$H\alpha$ emission. 
\cite{2001A&A...380..615C} presented an extensive data set of X~Per, 
covering  the period 1987 to 2001, 
and interpreted the HeI variability as a formation of a density wave
which moves towards the outer parts faster than the disc formation. 
It is likely that the four peaks profiles of X~Per
in 1995 represented a wave with eccentricity almost zero.

%  similar to the double disc  likely that 

Following
\cite{2001ApJ...546..455D}   %   Delgado-Marti et al. (2001) 
the periastron passages of the neutron star is expected at \\
MJD~$51353 (\pm 7)  + 250.3 (\pm 0.6)  \times E$. \\
The excitation of the eccentric wave could be connected with the periastron passage 
at about MJD~58361.  The periastron passage of the neutron star 
initiates (generates) a higher tidal wave on the stellar surface. Superposition of 
a high tide with another mechanism  (e.g. non-radial pulsations)    
could cause a mass ejection event from the star surface into the inner disc. 
A similar mass ejection was already discussed 
by  \cite{2014AJ....148..113L} in connection between the long term variability
of the $H\alpha$ line and the X-ray emission from the neutron star.     %  Li et al. 2014. 
%T_peri	51353	 ± 7
%Porb	250.3	 ± 0.6
The recent radiative transfer calculations on the structure of Be discs in coplanar circular binary 
systems suggest a V/R cycle every half orbital period 
\citep{2018MNRAS.473.3039P}.  % (Panoglou et al. 2018). 
We do not see signs of such modulation in X Per in our data, 
which covers 770 days, i.e. more than three orbital periods.

\section*{Conclusions}

We report 100 spectral observations of the Be/X-ray binary X~Per 
during the period December 2017 - January 2020. 
We study 
the evolution of the profiles of the emission lines 
$H\alpha$, $H\beta$, and $HeI6678$, 
which are formed in the Be circumstellar disc.
Their evolution suggests an eccentric wave in the circumstellar disc, for which we
find that:  (1) it spreads from inside out; (2) its velocity is $v_{wave} = 1.1 \pm 0.2$~
km~s$^{-1}$;
(3) the eccentricity of the eccentric wave is in the range $0.17 \le e_{wave} \le 0.41$. 

%To conclude we can say that 
The development of 
asymmetries in the inner and outer parts of the circumstellar disc
in the Be/X-ray binary X Per provides a laboratory to test 
the theoretical models of eccentric waves in the Be discs.    
                          % can be sometimes similar, sometimes different. 
We encourage high-resolution spectroscopic observations 
of this relatively  bright object 
(e.g. Echelle spectrographs on 2.0m class telescopes)
to monitor the evolution of the emission lines.

% This can help to understand the formation of large scale perturbations % and asymmetries
% in the circumstellar environment and their connection with accretion onto netron star.

\section*{Acknowledgments}

This work was supported by the  Bulgarian National Science Fund project  number K$\Pi$-06-H28/2 08.12.2018
"Binary stars with compact object".
The TIGRE telescope is a collaboration of the Hamburger
Sternwarte, the Universities of Hamburg, Guanajuato and Li\'ege.
UW acknowledges funding by DLR, project 50OR1701.
DM acknowledges partial support by grant RD-08-122/2020 from Shumen University.
This research has made use of  the MAXI data provided by RIKEN, JAXA, and the MAXI team 
and  observations from the AAVSO International Database 
contributed by observers worldwide.
We are very grateful to the anonymous referee for the very helpful and  constructive comments
on the original manuscript.

Data availability: The spectra are available upon request from the authors: rkz@astro.bas.bg, 
kstoyanov@astro.bas.bg. 

\bibliographystyle{mnras}
\bibliography{ref3}

% \bsp	% typesetting comme  nt
% \label{lastpage}	    			    
% \end{document}		    

%------------------------

% \newpage

\section{Appendix}
%---------------------------------------------
\begin{table}
  \centering
  \caption{ Journal of observations. In the table are given date of observation, observatory,
  exposure time, Julian day (24400000+). 
  }
   \begin{tabular}{lcrrcccccr} 
   \hline
   \\
      date-obs       &    Obs.   &  exp-time & JD & \\
                     &           &  [sec]    &    &  \\
2015-12-23T23:16     &   Rozhen  &  1800    &  57380.47503   &  \\
2015-12-23T23:48     &   Rozhen  &  600     &  57380.49672   &  \\
2015-12-24T20:51     &   Rozhen  &  600     &  57381.37398   &  \\
2015-12-26T20:01     &   Rozhen  &  600     &  57383.33892   &  \\
2015-12-27T20:56     &   Rozhen  &  300     &  57384.37750   &  \\
2016-01-30T19:10     &   Rozhen  &  300     &  57418.30099   &  \\
2016-09-23T01:42     &   Rozhen  &  600     &  57654.57407   &  \\
2016-09-23T01:42     &   Rozhen  &  600     &  57654.57407   &  \\
2016-12-11T20:58     &   Rozhen  &  1200    &  57734.37953   &  \\  
2017-03-17T18:58     &   Rozhen  &  1200    &  57830.28854   &  \\
2017-12-07T16:57     &   Rozhen  &  120     &  58095.21181   &  \\
2017-12-07T17:03     &   Rozhen  &  900     &  58095.21620   &  \\
2017-12-07T17:20     &   Rozhen  &  900     &  58095.22811   &  \\
2017-12-08T16:45     &   Rozhen  &  120     &  58096.20389   &  \\
2017-12-08T16:50     &   Rozhen  &  600     &  58096.20690   &  \\
2017-12-21T03:34     &   "LaLuz" &  240     &  58108.65399   &  \\
2017-12-22T00:53     &   "LaLuz" &  240     &  58109.54230   &  \\
2017-12-23T00:54     &   "LaLuz" &  127     &  58110.54298   &  \\  
2017-12-24T01:22     &   "LaLuz" &  240     &  58111.56198   &  \\
2017-12-25T01:26     &   "LaLuz" &  242     &  58112.56500   &  \\
2017-12-30T19:20     &   Rozhen  &  1200    &  58118.31067   &  \\
2018-01-01T17:23     &   Rozhen  &  1200    &  58120.22886   &  \\
2018-01-04T01:28     &   "LaLuz" &  900     &  58122.56566   &  \\
2018-01-04T03:56     &   "LaLuz" &  900     &  58122.66813   &  \\
2018-01-04T06:17     &   "LaLuz" &  900     &  58122.76651   &  \\
2018-01-05T01:21     &   "LaLuz" &  900     &  58123.56049   &  \\
2018-01-06T01:59     &   "LaLuz" &  900     &  58124.58737   &  \\  
2018-01-06T04:19     &   "LaLuz" &  900     &  58124.68461   &  \\
2018-01-06T06:39     &   "LaLuz" &  300     &  58124.78160   &  \\
2018-01-07T01:23     &   "LaLuz" &  900     &  58125.56225   &  \\
2018-01-07T03:49     &   "LaLuz" &  900     &  58125.66353   &  \\
2018-01-07T06:14     &   "LaLuz" &  900     &  58125.76423   &  \\
2018-01-26T17:09     &   Rozhen  &  120     &  58145.21745   &  \\
2018-01-26T17:12     &   Rozhen  &  180     &  58145.21951   &  \\
2018-01-26T17:16     &   Rozhen  &  900     &  58145.22261   &  \\
2018-01-26T17:32     &   Rozhen  &  1200    &  58145.23371   &  \\  
2018-02-01T17:09     &   Rozhen  &  1200    &  58151.21687   &  \\
2018-02-01T17:30     &   Rozhen  &  300     &  58151.23149   &  \\
2018-04-02T18:42     &   Rozhen  &  600     &  58211.27615   &  \\
2018-04-02T18:53     &   Rozhen  &  120     &  58211.28358   &  \\
2018-07-20T11:09     &   "LaLuz" &  600     &  58319.96179   &  \\
2018-08-02T11:20     &   "LaLuz" &  600     &  58332.97006   &  \\
2018-08-18T10:06     &   "LaLuz" &  600     &  58348.92028   &  \\
2018-08-19T01:10     &   Rozhen  &  120     &  58349.54794   &  \\
2018-08-19T01:14     &   Rozhen  &  900     &  58349.55114   &  \\  
2018-08-30T00:11     &   Rozhen  &  600     &  58360.50863   &  \\
2018-08-30T00:23     &   Rozhen  &  120     &  58360.51631   &  \\
2018-08-31T01:50     &   Rozhen  &  600     &  58361.57747   &  \\
2018-08-31T02:01     &   Rozhen  &  120     &  58361.58508   &  \\
2018-09-01T01:13     &   Rozhen  &  600     &  58362.55181   &  \\
2018-09-01T01:25     &   Rozhen  &  120     &  58362.55977   &  \\
2018-09-02T02:39     &   Rozhen  &  600     &  58363.61129   &  \\
2018-09-02T02:50     &   Rozhen  &  120     &  58363.61871   &  \\
2018-09-03T01:21     &   Rozhen  &  600     &  58364.55719   &  \\  
2018-09-03T01:31     &   Rozhen  &  120     &  58364.56449   &  \\
2018-09-05T08:54     &   "LaLuz" &  600     &  58366.87215   &  \\
2018-09-25T08:49     &   "LaLuz" &  600     &  58386.87070   &  \\
2018-10-10T07:16     &   "LaLuz" &  600     &  58401.80672   &  \\
2018-11-24T06:04     &   "LaLuz" &  600     &  58446.75842   &  \\
2018-11-25T22:30     &   Rozhen  &  180     &  58448.44317   &  \\
2018-12-06T07:11     &   "LaLuz" &  900     &  58458.80544   &  \\
2018-12-18T07:07     &   "LaLuz" &  900     &  58470.80209   &  \\
2018-12-20T20:54     &   Rozhen  &  120     &  58473.37615   &  \\  
2018-12-20T20:57     &   Rozhen  &  900     &  58473.37847   &  \\
  \\			    
 \hline 		    				     
 \end{tabular}  	    					      
 \label{tab.Jour}	    			    
 \end{table}		    			    
%--------------------------  ----------------
\begin{table}
  \centering
  \addtocounter{table}{-1}
  \caption{ Journal of observations (continuation). 
  }
   \begin{tabular}{lcrrrcccccr} 
   \hline
   \\
2018-12-21T18:18      &   Rozhen  &  120     &  58474.26792   &  \\
2018-12-21T18:21      &   Rozhen  &  900     &  58474.26989   &  \\
2018-12-27T22:31      &   Rozhen  &  600     &  58480.44336   &  \\
2018-12-27T22:42      &   Rozhen  &  120     &  58480.45117   &  \\
2018-12-30T07:12      &   "LaLuz" &  900     &  58482.80473   &  \\
2019-01-11T07:14      &   "LaLuz" &  900     &  58494.80544   &  \\
2019-01-18T20:16      &   Rozhen  &  60      &  58502.34774   &  \\
2019-01-18T20:19      &   Rozhen  &  120     &  58502.34988   &  \\  
2019-01-18T20:23      &   Rozhen  &  720     &  58502.35273   &  \\
2019-01-18T20:37      &   Rozhen  &  900     &  58502.36252   &  \\
2019-01-24T01:12      &   "LaLuz" &  900     &  58507.55335   &  \\
2019-02-07T01:37      &   "LaLuz" &  900     &  58521.56939   &  \\
2019-02-20T16:41      &   Rozhen  &  900     &  58535.19549   &  \\
2019-02-20T16:57      &   Rozhen  &  120     &  58535.20681   &  \\
2019-02-21T17:38      &   Rozhen  &  1200    &  58536.23481   &  \\
2019-02-21T17:58      &   Rozhen  &  120     &  58536.24923   &  \\
2019-03-02T01:51      &   "LaLuz" &  1125    &  58544.57689   &  \\  
2019-03-17T18:50      &   Rozhen  &  600     &  58560.28307   &  \\
2019-03-17T19:02      &   Rozhen  &  120     &  58560.29132   &  \\
2019-03-23T19:10      &   Rozhen  &  120     &  58566.29626   &  \\
2019-03-23T19:14      &   Rozhen  &  900     &  58566.29891   &  \\
2019-03-25T18:19      &   Rozhen  &  120     &  58568.26056   &  \\
2019-03-25T18:22      &   Rozhen  &  900     &  58568.26288   &  \\
2019-07-18T00:30      &   Rozhen  &  600     &  58682.51763   &  \\
2019-07-18T00:41      &   Rozhen  &  120     &  58682.52516   &  \\
2019-07-19T00:51      &   Rozhen  &  600     &  58683.53222   &  \\  
2019-07-19T01:01      &   Rozhen  &  120     &  58683.53957   &  \\
2019-08-19T00:43      &   Rozhen  &  600     &  58714.52967   &  \\
2019-08-19T00:54      &   Rozhen  &  1200    &  58714.53737   &  \\
2019-08-22T23:06      &   Rozhen  &  600     &  58718.46281   &  \\
2019-08-22T23:18      &   Rozhen  &  120     &  58718.47082   &  \\
2019-09-15T00:36      &   Rozhen  &  900     &  58741.52743   &  \\
2019-09-15T00:53      &   Rozhen  &  120     &  58741.53915   &  \\
2019-11-15T22:43      &   Rozhen  &  900     &  58803.45246   &  \\
2019-11-15T22:43      &   Rozhen  &  900     &  58803.45246   &  \\  
2019-12-05T20:30      &   Rozhen  &  600     &  58823.35991   &  \\
2019-12-05T20:41      &   Rozhen  &  120     &  58823.36759   &  \\
2019-12-06T19:02      &   Rozhen  &  600     &  58824.29886   &  \\
2019-12-06T19:13      &   Rozhen  &  120     &  58824.30628   &  \\
2020-01-03T17:55      &   Rozhen  &  600     &  58852.25124   &  \\
2020-01-03T18:06      &   Rozhen  &  120     &  58852.25892   &  \\
2020-01-07T16:51      &   Rozhen  &  1200    &  58856.20631   &  \\
2020-01-07T17:12      &   Rozhen  &  3600    &  58856.22133   &  \\
2020-01-10T17:17      &   Rozhen  &  120     &  58859.22428   &  \\  
2020-01-10T17:23      &   Rozhen  &  900     &  58859.22862   &  \\
2020-01-16T19:51      &   Rozhen  &  900     &  58865.33101   &  \\
2020-01-16T20:08      &   Rozhen  &  120     &  58865.34259   &  \\
  \\			    
 \hline 		    				     
 \end{tabular}  	    					      
 \end{table}		    			    
%--------------------------  ----------------
			    
%---------------------------------------------
\begin{table*}
  \centering
  \caption{ Parameters of the $H\alpha$ emission line. In the table are given 
  Julian day (2400000+), equivalent width, intensity of the violet and red peaks, 
  distance between the peaks, radial velocities of the  peaks.  
  }
   \begin{tabular}{lrrrrrrr} 
   \hline
  \\
JD           &   $W\alpha$  & $I_B$   &  $I_R$   & $\Delta V$  &    V        &    V           & \\
             &              &         &          & [km s$^{-1}$] & [km s$^{-1}$] &   [km s$^{-1}$]  & \\
\\
58095.21084  &     -24.96   & 5.530   &  5.547   &  119.8  &	   -72.3   &  48.2 & \\
58095.21523  &     -24.09   & 5.316   &  5.252   &  121.0  &	   -72.9   &  48.0 & \\
58095.22714  &     -23.61   & 5.365   &  5.316   &  119.7  &	   -72.7   &  47.7 & \\
58096.20296  &     -24.43   & 5.581   &  5.598   &  119.0  &	   -72.7   &  47.6 & \\
58096.20597  &     -24.22   & 5.521   &  5.422   &  121.6  &	   -73.3   &  49.3 & \\
58118.31064  &     -22.61   & 4.954   &  5.006   &  123.7  &	   -77.5   &  45.9 & \\
58120.22892  &     -20.32   & 4.839   &  4.875   &  120.8  &	   -76.0   &  44.0 & \\
58145.21851  &     -18.61   & 4.162   &  4.709   &  108.0  &	   -74.5   &  33.5 & \\
58145.22057  &     -18.67   & 4.146   &  4.755   &  108.4  &	   -77.0   &  33.9 & \\
58145.22368  &     -18.25   & 4.141   &  4.741   &  110.3  &	   -76.5   &  33.7 & \\
58145.23478  &     -18.55   & 4.146   &  4.619   &  109.6  &	   -76.2   &  33.1 & \\
58151.21816  &     -18.77   & 4.076   &  4.859   &  116.2  &	   -85.1   &  32.5 & \\
58151.23277  &     -17.68   & 4.068   &  4.881   &  116.1  &	   -82.5   &  32.3 & \\
58211.27855  &     -16.35   & 3.519   &  4.480   &  106.6  &	   -89.0   &  17.6 & \\
58211.28598  &     -15.92   & 3.628   &  4.511   &  102.4  &	   -86.8   &  17.9 & \\
58349.54619  &     -18.81   & 3.440   &  5.197   &  105.3  &	   -98.0   &  12.4 & \\
58349.54938  &     -17.66   & 3.409   &  5.159   &  98.0   &	   -98.7   &  10.3 & \\
58360.50660  &     -18.42   & 3.278   &  5.383   &  100.7  &	   -93.8   &  9.2  & \\
58360.51428  &     -18.84   & 3.309   &  5.345   &  105.8  &	   -96.8   &  10.8 & \\
58361.57541  &     -17.77   & 3.139   &  5.173   &  97.8   &	   -90.0   &  3.2  & \\
58361.58303  &     -18.42   & 3.158   &  5.258   &  87.9   &	   -89.5   &  2.9  & \\
58362.54974  &     -18.92   & 3.247   &  5.363   &  94.8   &	   -89.0   &  5.3  & \\
58362.55770  &     -18.49   & 3.245   &  5.343   &  100.8  &	   -92.8   &  3.9  & \\
58363.60919  &     -19.45   & 3.292   &  5.438   &  98.5   &	   -94.8   &  1.5  & \\
58363.61661  &     -18.86   & 3.284   &  5.417   &  95.2   &	   -93.8   &  2.5  & \\
58364.55507  &     -18.88   & 3.304   &  5.386   &  95.2   &	   -96.9   &  2.9  & \\
58364.56238  &     -19.03   & 3.293   &  5.419   &  99.9   &	   -93.5   &  0.8  & \\
58448.44176  &     -21.17   & 3.371   &  5.767   &  115.9  &	   -115.0  &  2.1  & \\
58473.37570  &     -19.99   & 3.663   &  5.509   &  108.2  &	   -103.3  &  7.7  & \\
58473.37802  &     -20.87   & 3.607   &  5.397   &  115.2  &	   -106.1  &  6.3  & \\
58474.26750  &     -20.30   & 3.606   &  5.448   &  110.6  &	   -101.9  &  9.2  & \\
58474.26947  &     -20.24   & 3.507   &  5.400   &  112.9  &	   -105.6  &  5.8  & \\
58480.44321  &     -21.40   & 3.710   &  5.538   &  112.0  &	   -106.1  &  8.1  & \\
58480.45101  &     -21.28   & 3.802   &  5.691   &  110.4  &	   -106.5  &  8.8  & \\
58502.34850  &     -20.51   & 3.822   &  5.251   &  105.8  &	   -99.7   &  11.6 & \\
58502.35064  &     -20.05   & 3.836   &  5.252   &  109.7  &	   -104.7  &  12.1 & \\
58502.35348  &     -20.54   & 3.809   &  5.196   &  109.3  &	   -103.1  &  12.0 & \\
58502.36328  &     -20.25   & 3.833   &  5.222   &  112.0  &	   -102.5  &  11.7 & \\
58535.19736  &     -19.42   & 3.714   &  5.307   &  114.6  &	   -94.8   &  21.8 & \\
58535.20868  &     -20.20   & 3.747   &  5.353   &  115.9  &	   -94.4   &  20.4 & \\
58536.23671  &     -20.01   & 3.678   &  5.219   &  116.5  &	   -97.8   &  21.7 & \\
58536.25113  &     -19.68   & 3.743   &  5.286   &  117.6  &	   -98.7   &  20.8 & \\
58560.28540  &     -20.20   & 3.905   &  4.708   &  108.3  &	   -87.3   &  21.9 & \\
58560.29365  &     -19.43   & 3.893   &  4.837   &  106.9  &	   -85.9   &  21.0 & \\
58566.29864  &     -21.40   & 4.367   &  4.845   &  108.3  &	   -87.6   &  26.7 & \\
58566.30129  &     -20.18   & 4.117   &  4.788   &  110.6  &	   -86.3   &  25.8 & \\
58568.26295  &     -19.94   & 4.175   &  4.693   &  107.3  &	   -82.3   &  25.3 & \\
58568.26527  &     -20.45   & 4.093   &  4.686   &  108.3  &	   -81.8   &  25.4 & \\
58108.65545  &     -23.37   & 5.518   &  5.257   &  125.3  &	   -69.8   &  56.3 & \\
58109.54377  &     -23.51   & 5.421   &  5.273   &  121.6  &	   -66.1   &  58.4 & \\
58110.54374  &     -24.88   & 5.477   &  5.217   &  123.8  &	   -69.0   &  55.1 & \\
58111.56345  &     -26.02   & 5.385   &  5.244   &  125.2  &	   -72.0   &  50.6 & \\
58112.56647  &     -23.20   & 5.046   &  5.056   &  123.9  &	   -65.5   &  57.3 & \\
58122.57100  &     -20.95   & 4.840   &  4.850   &  124.3  &	   -72.3   &  51.6 & \\
58122.67347  &     -21.42   & 4.890   &  4.931   &  123.4  &	   -73.8   &  50.4 & \\
58122.77185  &     -21.03   & 4.859   &  4.884   &  124.6  &	   -74.0   &  50.4 & \\
58123.56582  &     -22.24   & 5.003   &  5.008   &  125.2  &	   -72.7   &  51.3 & \\
  \\			    
 \hline 		    				     
 \end{tabular}  	    					      
 \label{tab.Ha}	    			    
 \end{table*}		    			    
%--------------------------  ----------------
\begin{table*}
  \centering
  \addtocounter{table}{-1}
  \caption{ continuation. 
  }
   \begin{tabular}{lrrrrrrr} 
   \hline
   \\
58124.59271  &     -22.08   & 4.983   &  4.921   &  122.4  &	   -74.1   &  49.0 & \\
58124.68994  &     -20.87   & 4.933   &  4.892   &  122.4  &	   -74.8   &  47.8 & \\
58124.78338  &     -20.99   & 4.921   &  4.856   &  122.6  &	   -75.1   &  48.5 & \\
58125.56759  &     -20.60   & 4.849   &  4.829   &  123.2  &	   -72.0   &  50.9 & \\
58125.66887  &     -20.71   & 4.888   &  4.872   &  122.0  &	   -72.9   &  49.2 & \\
58125.76956  &     -21.29   & 4.902   &  4.887   &  122.2  &	   -73.7   &  49.6 & \\
58319.96541  &     -16.63   & 3.143   &  4.770   &  107.5  &	   -95.4   &  18.2 & \\
58332.97367  &     -16.86   & 3.144   &  5.053   &  107.9  &	   -96.7   &  16.2 & \\
58348.92388  &     -19.44   & 3.499   &  5.272   &  100.4  &	   -96.5   &  12.7 & \\
58366.87575  &     -20.47   & 3.436   &  5.451   &  100.3  &	   -93.3   &  12.4 & \\
58386.87432  &     -20.88   & 3.508   &  5.680   &  98.9   &	   -102.3  &  6.1  & \\
58401.81034  &     -21.58   & 3.908   &  5.515   &  100.3  &	   -96.5   &  8.1  & \\
58446.76209  &     -20.95   & 3.400   &  5.717   &  120.5  &	   -112.6  &  8.3  & \\
58458.81081  &     -19.76   & 3.172   &  5.571   &  117.9  &	   -116.1  &  12.3 & \\
58470.80746  &     -20.75   & 3.496   &  5.368   &  114.9  &	   -108.5  &  13.3 & \\
58482.81011  &     -20.85   & 3.744   &  5.403   &  105.9  &	   -99.0   &  10.7 & \\
58494.81083  &     -21.35   & 3.815   &  5.372   &  90.4   &	   -85.8   &  10.6 & \\
58507.55874  &     -20.20   & 3.730   &  5.266   &  102.4  &	   -95.2   &  16.5 & \\
58521.57478  &     -20.01   & 3.636   &  5.246   &  116.7  &	   -98.1   &  20.6 & \\
58544.58365  &     -20.18   & 3.939   &  5.038   &  109.6  &	   -84.3   &  31.4 & \\
58682.52460  &     -16.63   & 4.196   &  3.755   &  110.5  &	   -82.3   &  26.0 & \\
58682.52940  &     -17.08   & ---     &  ---     &  ---    &	    ---    &  ---  & \\
58683.53920  &     -17.11   & 4.182   &  3.864   &  112.7  &	   -83.0   &  24.2 & \\
58683.54370  &     -16.64   & 4.261   &  3.931   &  109.1  &	   -73.7   &  21.7 & \\
58714.53390  &     -15.77   & 3.998   &  3.579   &  103.3  &	   -75.1   &  25.5 & \\
58714.54510  &     -15.39   & 4.002   &  3.633   &  108.1  &	   -75.1   &  28.5 & \\
58718.46660  &     -15.81   & 3.925   &  3.686   &  110.3  &	   -75.6   &  34.2 & \\
58718.47190  &     -15.64   & 3.951   &  3.740   &  110.3  &	   -76.4   &  36.0 & \\
58741.53080  &     -14.26   & 4.135   &  3.377   &  106.1  &	   -68.5   &  34.3 & \\
58741.53800  &     -14.31   & 4.158   &  3.432   &  102.1  &	   -67.5   &  28.7 & \\
58803.45210  &     -12.39   & 3.829   &  2.898   &  112.9  &	   -55.4   &  57.2 & \\
58803.45210  &     -12.49   & 3.835   &  2.916   &  111.1  &	   -55.7   &  56.4 & \\
58823.35790  &     -11.58   & 3.637   &  2.550   &  107.7  &	   -50.6   &  59.7 & \\
58823.36280  &     -11.03   & 3.625   &  2.548   &  110.5  &	   -49.6   &  63.9 & \\
58824.29680  &     -11.66   & 3.621   &  2.482   &  109.1  &	   -55.1   &  58.6 & \\
58824.30150  &     -10.92   & 3.618   &  2.481   &  113.3  &	   -55.8   &  59.7 & \\
58852.25040  &     -10.15   & 3.294   &  2.374   &  119.2  &	   -58.2   &  56.4 & \\
58852.25530  &     -10.26   & 3.339   &  2.398   &  114.4  &	   -58.1   &  58.1 & \\
58856.20920  &     -10.51   & 3.251   &  2.395   &  106.8  &	   -56.3   &  54.8 & \\
58856.23810  &     -10.32   & 3.179   &  2.347   &  110.5  &	   -58.6   &  55.1 & \\
58859.22110  &     -10.33   & 3.176   &  2.372   &  125.3  &	   -58.7   &  67.4 & \\
58859.23000  &     -10.06   & 3.153   &  2.362   &  127.7  &	   -58.8   &  65.6 & \\
58865.33290  &     -10.20   & 3.161   &  2.370   &  120.8  &	   -59.7   &  63.1 & \\
58865.33990  &     -10.19   & 3.231   &  2.414   &  119.4  &	   -60.4   &  62.9 & \\  
  \\			    
 \hline 		    				     
 \end{tabular}  	    					      	    			    
 \end{table*}		    			    
%--------------------------  ----------------

%---------------------------------------------
\begin{table*}
  \centering
  \caption{ Parameters of the $H\beta$ emission line. In the table are given 
  Julian day (2400000+), equivalent width, intensity of the violet and red peaks, distance between the peaks.  
  }
   \begin{tabular}{lrrrrrrrcc} 
   \hline
   \\
58095.21084   &    -4.357  &  2.094  &   2.167  &   150.5  & 	-87.3  &   61.7   &   ---   & \\
58095.21523   &    -4.140  &  2.092  &   2.115  &   150.9  & 	-86.5  &   64.2   &   ---   & \\
58095.22714   &    -4.172  &  2.088  &   2.116  &   151.0  & 	-87.3  &   63.9   &   ---   & \\
58096.20296   &    -4.132  &  2.161  &   2.131  &   148.9  & 	-85.8  &   63.3   &   ---   & \\
58096.20597   &    -4.184  &  2.122  &   2.115  &   148.8  & 	-88.1  &   61.7   &   ---   & \\
58118.31064   &    -3.699  &  1.965  &   2.094  &   148.1  & 	-91.9  &   58.7   &   ---   & \\
58120.22892   &    -3.814  &  1.944  &   2.106  &   143.9  & 	-87.9  &   56.7   &   ---   & \\
58145.21851   &    -3.262  &  1.775  &   2.039  &   141.5  & 	-78.7  &   65.7   &   ---   & \\
58145.22057   &    -3.315  &  1.763  &   2.048  &   137.0  & 	-78.1  &   62.8   &   ---   & \\
58145.22368   &    -3.322  &  1.754  &   2.015  &   140.1  & 	-78.9  &   63.3   &   ---   & \\
58145.23478   &    -3.360  &  1.762  &   2.023  &   140.3  & 	-80.2  &   62.5   &   ---   & \\
58151.21816   &    -3.259  &  1.725  &   2.074  &   138.0  & 	-102.0 &   40.0   &   ---   & \\
58151.23277   &    -3.509  &  1.747  &   2.088  &   141.8  & 	-107.1 &   40.0   &   ---   & \\
58211.27855   &    -2.544  &  1.548  &   1.918  &   106.5  & 	-84.4  &   28.8   &   ---   & \\
58211.28598   &    -2.683  &  1.545  &   1.936  &   127.6  & 	-97.2  &   28.0   &   ---   & \\
58349.54619   &    -3.123  &  1.520  &   2.095  &   125.0  & 	-119.1 &   19.3   &   ---   & \\
58349.54938   &    -3.150  &  1.505  &   2.018  &   130.0  & 	-113.1 &   21.7   &   ---   & \\
58360.50660   &    -2.963  &  1.449  &   2.035  &   132.7  & 	-120.0 &   19.6   &   ---   & \\
58360.51428   &    -3.062  &  1.469  &   2.059  &   131.5  & 	-115.7 &   18.7   &   ---   & \\
58361.57541   &    -3.034  &  1.413  &   2.021  &   131.2  & 	-116.9 &   19.6   &   ---   & \\
58361.58303   &    -3.096  &  1.437  &   2.040  &   135.7  & 	-111.8 &   20.4   &   ---   & \\
58362.54974   &    -2.815  &  1.439  &   2.072  &   137.4  & 	-116.8 &   17.9   &   ---   & \\
58362.55770   &    -3.151  &  1.447  &   2.062  &   139.1  & 	-124.8 &   19.3   &   ---   & \\
58363.60919   &    -3.133  &  1.434  &   2.066  &   140.0  & 	-121.7 &   18.4   &   ---   & \\
58363.61661   &    -2.993  &  1.437  &   2.068  &   130.8  & 	-117.9 &   16.0   &   ---   & \\
58364.55507   &    -3.103  &  1.460  &   2.038  &   141.3  & 	-155.3 &   -18.6  &   ---   & \\
58364.56238   &    -3.072  &  1.475  &   2.023  &   144.1  & 	-153.9 &   -16.1  &   ---   & \\
58448.44176   &    -3.263  &  1.552  &   2.021  &   163.8  & 	-155.9 &   10.1   &   ---   & \\
58473.37570   &    -3.328  &  1.576  &   2.012  &   147.7  & 	-122.6 &   25.2   &   ---   & \\
58473.37802   &    -3.450  &  1.548  &   1.985  &   147.8  & 	-123.4 &   22.2   &   ---   & \\
58474.26750   &    -3.288  &  1.535  &   2.106  &   158.2  & 	-130.1 &   25.0   &   ---   & \\
58474.26947   &    -3.219  &  1.515  &   2.010  &   148.9  & 	-124.8 &   23.5   &   ---   & \\
58480.44321   &    -3.179  &  1.549  &   2.005  &   153.6  & 	-127.9 &   21.5   &   ---   & \\
58480.45101   &    -3.078  &  1.549  &   2.026  &   148.9  & 	-126.9 &   20.6   &   ---   & \\
58502.34850   &    -3.392  &  1.597  &   1.933  &   145.1  & 	-99.0  &   40.1   &   ---   & \\
58502.35064   &    -3.229  &  1.590  &   1.910  &   150.2  & 	-107.6 &   39.4   &   ---   & \\
58502.35348   &    -3.347  &  1.601  &   1.932  &   149.4  & 	-108.9 &   41.1   &   ---   & \\
58502.36328   &    -3.307  &  1.588  &   1.923  &   152.4  & 	-109.7 &   40.1   &   ---   & \\
58535.19736   &    -3.198  &  1.569  &   1.866  &   150.0  & 	-122.2 &   27.5   &   ---   & \\
58535.20868   &    -3.131  &  1.590  &   1.941  &   152.8  & 	-125.8 &   26.1   &   ---   & \\
58536.23671   &    -2.858  &  1.574  &   1.875  &   158.9  & 	-128.8 &   29.1   &   ---   & \\
58536.25113   &    -2.988  &  1.601  &   1.897  &   161.1  & 	-127.4 &   29.4   &   ---   & \\
58560.28540   &    -2.870  &  1.629  &   1.726  &   147.6  & 	-118.7 &   29.5   &   ---   & \\
58560.29365   &    -2.885  &  1.661  &   1.739  &   145.0  & 	-118.6 &   27.7   &   ---   & \\
58566.29864   &    -3.198  &  1.772  &   1.809  &   146.9  & 	-117.2 &   28.4   &   ---   & \\
58566.30129   &    -2.738  &  1.672  &   1.691  &   ---    & 	---    &   ---    &   ---   & \\
58568.26295   &    -2.934  &  1.706  &   1.732  &   148.7  & 	-115.8 &   25.1   &   ---   & \\
58568.26527   &    -3.073  &  1.698  &   1.707  &   146.0  & 	-115.5 &   30.4   &   ---   & \\
58108.65545   &    -4.003  &  1.969  &   2.157  &   152.7  & 	-84.4  &   68.1   &   ---   & \\
58109.54377   &    -3.924  &  1.971  &   2.126  &   145.7  & 	-83.1  &   62.5   &   ---   & \\
58110.54374   &    -4.017  &  2.109  &   2.273  &   145.2  & 	---    &   ---    &   ---   & \\
58111.56345   &    -4.062  &  1.999  &   2.206  &   149.0  & 	-88.7  &   63.9   &   ---   & \\
58112.56647   &    -3.892  &  2.006  &   2.168  &   147.0  & 	-84.4  &   63.1   &   ---   & \\
58122.57100   &    -3.621  &  1.835  &   2.059  &   148.4  & 	-94.9  &   58.0   &   ---   & \\
58122.67347   &    -3.521  &  1.892  &   2.053  &   149.8  & 	-91.5  &   56.0   &   ---   & \\
58122.77185   &    -3.525  &  1.845  &   2.039  &   150.0  & 	-94.5  &   57.7   &   ---   & \\
58123.56582   &    -3.613  &  1.880  &   2.062  &   148.7  & 	-90.5  &   57.6   &   ---   & \\
58124.59271   &    -3.615  &  1.899  &   2.066  &   146.3  & 	-87.6  &   58.1   &   ---   & \\
58124.68994   &    -3.488  &  1.868  &   2.023  &   145.8  & 	-89.9  &   54.9   &   ---   & \\
58124.78338   &    -3.548  &  1.855  &   2.047  &   143.8  & 	-86.6  &   55.0   &   ---   & \\
  \\			    
 \hline 		    				     
 \end{tabular}  	    					      
 \label{tab.Hb}	    			    
 \end{table*}		    			    
%--------------------------  ----------------
\begin{table*}
  \centering
  \addtocounter{table}{-1}
  \caption{ continuation. 
  }
   \begin{tabular}{lrrrrrrrcc} 
   \hline
   \\
58125.56759   &    -3.582  &  1.871  &   2.059  &   140.7  & 	-89.7  &   53.3   &   ---   & \\
58125.66887   &    -3.569  &  1.917  &   2.035  &   142.8  & 	-89.2  &   54.6   &   ---   & \\
58125.76956   &    -3.541  &  1.859  &   2.021  &   143.0  & 	-88.7  &   55.5   &   ---   & \\
58319.96541   &    -2.779  &  1.442  &   1.966  &   120.2  & 	-97.5  &   26.2   &   ---   & \\
58332.97367   &    -2.733  &  1.450  &   1.980  &   126.0  & 	---    &   ---    &   ---   & \\
58348.92388   &    -2.979  &  1.536  &   2.031  &   127.2  & 	-106.0 &   21.4   &   ---   & \\
58366.87575   &    -3.002  &  1.505  &   2.076  &   124.7  & 	-106.5 &   18.9   &   ---   & \\
58386.87432   &    -3.309  &  1.474  &   2.056  &   144.2  & 	-130.4 &   18.4   &   ---   & \\
58401.81034   &    -3.424  &  1.598  &   2.047  &   139.0  & 	-124.3 &   17.0   &   ---   & \\
58446.76209   &    -3.257  &  1.578  &   2.034  &   166.9  & 	-152.5 &   16.2   &   ---   & \\
58458.81081   &    -3.001  &  1.509  &   2.006  &   167.5  & 	-147.9 &   20.1   &   ---   & \\
58470.80746   &    -3.126  &  1.541  &   1.964  &   152.3  & 	-128.5 &   23.2   &   ---   & \\
58482.81011   &    -3.067  &  1.556  &   1.911  &   141.6  & 	-121.9 &   21.7   &   ---   & \\
58494.81083   &    -3.104  &  1.523  &   1.956  &   146.7  & 	-126.4 &   21.1   &   ---   & \\
58507.55874   &    -3.032  &  1.565  &   1.852  &   156.7  & 	-127.8 &   27.3   &   ---   & \\
58521.57478   &    -2.855  &  1.567  &   1.842  &   157.9  & 	-132.4 &   27.8   &   ---   & \\
58544.58365   &    -2.764  &  1.584  &   1.773  &   155.7  & 	-114.4 &   38.8   &   ---   & \\
58682.52460   &    -2.246  &  1.630  &   1.400  &   117.1  & 	-98.3  &   23.3   &   220.0 & \\
58682.52940   &    ---     &  ----   &   ---    &   ---    & 	---    &   ---    &   ---   & \\
58683.53920   &    -2.309  &  1.598  &   1.377  &   113.2  & 	-100.5 &   23.8   &   206.6 & \\
58683.54370   &    -2.355  &  1.604  &   1.400  &   109.8  & 	-100.4 &   17.1   &   196.9 & \\
58714.53390   &    -2.100  &  1.619  &   1.357  &   123.3  & 	-90.8  &   24.9   &   202.2 & \\
58714.54510   &    -2.015  &  1.603  &   1.358  &   119.9  & 	-91.5  &   31.5   &   197.5 & \\
58718.46660   &    -2.039  &  1.608  &   1.389  &   122.0  & 	-89.8  &   34.2   &   197.7 & \\
58718.47190   &    -2.009  &  1.598  &   1.433  &   122.2  & 	-89.7  &   31.0   &   191.5 & \\
58741.53080   &    -1.896  &  1.615  &   1.315  &   125.3  & 	-81.0  &   47.2   &   207.0 & \\
58741.53800   &    -2.019  &  1.661  &   1.324  &   141.8  & 	-81.2  &   24.8   &   194.1 & \\
58803.45210   &    -1.675  &  1.536  &   1.261  &   131.4  & 	-70.8  &   65.4   &   170.1 & \\
58803.46410   &    -1.594  &  1.518  &   1.248  &   130.0  & 	-70.2  &   61.8   &   163.1 & \\
58823.35790   &    -1.571  &  1.502  &   1.215  &   128.1  & 	-65.7  &   65.9   &   162.5 & \\
58823.36280   &    -1.628  &  1.509  &   1.251  &   126.2  & 	-65.4  &   63.7   &   166.7 & \\
58824.29680   &    -1.456  &  1.497  &   1.202  &   128.7  & 	-68.7  &   65.2   &   153.4 & \\
58824.30150   &    -1.507  &  1.479  &   1.224  &   127.3  & 	-68.8  &   60.1   &   151.3 & \\
58852.25040   &    -1.438  &  1.439  &   1.119  &   149.0  & 	-68.7  &   89.0   &   ---   & \\
58852.25530   &    -1.473  &  1.467  &   1.225  &   148.9  & 	-69.1  &   96.8   &   ---   & \\
58856.20920   &    -1.498  &  1.415  &   1.242  &   155.8  & 	-67.9  &   88.3   &   ---   & \\
58856.23810   &    -1.481  &  1.429  &   1.221  &   147.2  & 	-68.4  &   86.5   &   ---   & \\
58859.22110   &    -1.475  &  1.422  &   1.224  &   152.3  & 	-68.7  &   89.3   &   ---   & \\
58859.23000   &    -1.427  &  1.421  &   1.210  &   148.7  & 	-68.6  &   78.7   &   ---   & \\
58865.33290   &    -1.427  &  1.434  &   1.244  &   173.3  & 	-67.6  &   96.1   &   ---   & \\
58865.33990   &    -1.366  &  1.431  &   1.259  &   165.5  & 	-69.4  &   87.3   &   ---   & \\
  \\			    
 \hline 		    				     
 \end{tabular}  	    					      	    			    
 \end{table*}		    			    
%---------------------------------------

%---------------------------------------------
\begin{table*}
  \centering
  \caption{ Parameters of the HeI~6678 emission line. In the table are given 
  Julian day (2400000+), equivalent width, intensity of the violet and red peaks, 
  distance between the peaks, radial velocities of the  peaks.  
  }
   \begin{tabular}{lrrrrrrrccc} 
   \hline
  \\
JD           &   $W\alpha$  & $I_B$   &  $I_R$   & $\Delta V$  &    V        &    V           &    V	       & \\
             &              &         &          & [km s$^{-1}$] & [km s$^{-1}$] &   [km s$^{-1}$]  &   [km s$^{-1}$]  & \\
\\	        									      
58095.21084  &       -2.004  &  1.363 &    1.346  &   181.9   &  -98.0   &  81.7  &            &  \\ 
58095.21523  &       -1.908  &  1.334 &    1.323  &   182.5   &  -95.5   &  84.2  &            &  \\
58095.22714  &       -1.917  &  1.327 &    1.325  &   178.9   &  -100.2  &  82.7  &            &  \\
58096.20296  &       -2.030  &  1.344 &    1.363  &   178.3   &  -101.0  &  81.0  &            &  \\
58096.20597  &       -1.973  &  1.340 &    1.330  &   179.3   &  -98.3   &  84.8  &            &  \\
58118.31064  &       -1.564  &  1.278 &    1.301  &   180.9   &  -111.3  &  71.2  &            &  \\
58120.22892  &       -1.470  &  1.264 &    1.272  &   169.3   &  -102.8  &  67.8  &            &  \\
58145.21851  &       -1.118  &  1.185 &    1.240  &   159.1   &  -105.6  &  52.8  &            &  \\
58145.22057  &       -1.229  &  1.179 &    1.251  &   154.6   &  -99.8   &  53.6  &            &  \\
58145.22368  &       -1.193  &  1.174 &    1.246  &   155.1   &  -106.6  &  53.1  &            &  \\
58145.23478  &       -1.202  &  1.183 &    1.242  &   164.2   &  -111.5  &  54.2  &            &  \\
58151.21816  &       -1.180  &  1.178 &    1.263  &   165.3   &  -116.1  &  49.5  &            &  \\
58151.23277  &       -1.073  &  1.193 &    1.271  &   168.0   &  -115.8  &  52.8  &            &  \\
58211.27855  &       -0.865  &  1.107 &    1.206  &   166.8   &  -127.2  &  43.1  &            &  \\
58211.28598  &       -0.922  &  1.114 &    1.209  &   ---     &   ---	 &  ---   &            &  \\
58349.54619  &       -0.904  &  1.110 &    1.210  &   201.8   &  -164.5  &  34.3  &            &  \\
58349.54938  &       -0.816  &  1.100 &    1.195  &   202.8   &  -169.8  &  35.3  &            &  \\
58360.50660  &       -0.932  &  1.109 &    1.196  &   212.0   &  -181.8  &  29.9  &            &  \\
58360.51428  &       -0.907  &  1.113 &    1.209  &   190.3   &  -165.4  &  26.0  &            &  \\
58361.57541  &       -0.939  &  1.109 &    1.212  &   211.7   &  -185.9  &  29.1  &            &  \\
58361.58303  &       -0.911  &  1.106 &    1.218  &   200.5   &  -173.1  &  28.8  &            &  \\
58362.54974  &       -0.927  &  1.095 &    1.214  &   201.2   &  -167.9  &  30.2  &            &  \\
58362.55770  &       -0.950  &  1.103 &    1.205  &   196.0   &  -161.1  &  31.5  &            &  \\
58363.60919  &       -1.044  &  1.099 &    1.219  &   209.0   &  -179.9  &  29.8  &            &  \\
58363.61661  &       -1.089  &  1.107 &    1.231  &   204.9   &  -172.4  &  29.9  &            &  \\
58364.55507  &       -1.112  &  1.107 &    1.200  &   209.7   &  -173.5  &  28.8  &            &  \\
58364.56238  &       -1.104  &  1.121 &    1.205  &   201.8   &  -172.6  &  30.6  &            &  \\
58448.44176  &       -1.586  &  1.197 &    1.243  &   206.9   &  -182.7  &  25.4  &            &  \\
58473.37570  &       -1.749  &  1.201 &    1.244  &   217.6   &  -189.1  &  32.5  &            &  \\
58473.37802  &       -1.665  &  1.178 &    1.210  &   211.5   &  -188.1  &  27.2  &    199.1   &  \\
58474.26750  &       -1.540  &  1.218 &    1.262  &   212.2   &  -185.4  &  25.3  &    ---     &  \\
58474.26947  &       -1.573  &  1.186 &    1.212  &   218.3   &  -190.9  &  26.5  &    187.0   &  \\
58480.44321  &       -1.653  &  1.214 &    1.227  &   218.8   &  -190.9  &  28.0  &    163.3   &  \\
58480.45101  &       -1.942  &  1.265 &    1.297  &   205.2   &  -185.5  &  26.1  &    ---     &  \\
58502.34850  &       -1.550  &  1.233 &    1.216  &   199.9   &  -176.6  &  22.1  &    ---     &  \\
58502.35064  &       -1.580  &  1.208 &    1.194  &   201.7   &  -171.3  &  28.6  &    ---     &  \\
58502.35348  &       -1.475  &  1.212 &    1.197  &   198.4   &  -166.7  &  28.0  &    205.8   &  \\
58502.36328  &       -1.591  &  1.211 &    1.193  &   201.7   &  -172.5  &  27.5  &    206.8   &  \\
58535.19736  &       -1.505  &  1.205 &    1.178  &   191.2   &  -157.2  &  32.2  &    231.0   &  \\
58535.20868  &       -1.536  &  1.217 &    1.195  &   192.6   &  -162.1  &  34.6  &    204.6   &  \\
58536.23671  &       -1.505  &  1.205 &    1.171  &   180.2   &  -149.2  &  34.5  &    229.0   &  \\
58536.25113  &       -1.522  &  1.221 &    1.183  &   185.4   &  -152.7  &  31.8  &    234.8   &  \\
58560.28540  &       -1.427  &  1.216 &    1.138  &   177.4   &  -144.8  &  30.0  &    246.5   &  \\
58560.29365  &       -1.520  &  1.232 &    1.130  &   176.2   &  -152.9  &  28.7  &    223.3   &  \\
58566.29864  &       -1.648  &  1.242 &    1.175  &   -1.0    &   ---	 &  ---   &    ---     &  \\
58566.30129  &       -1.470  &  1.220 &    1.141  &   160.2   &  -144.8  &  23.5  &    245.2   &  \\
58568.26295  &       -1.486  &  1.229 &    1.159  &   141.9   &  -138.9  &  11.1  &    250.1   &  \\
58568.26527  &       -1.375  &  1.224 &    1.142  &   162.2   &  -133.3  &  25.5  &    240.9   &  \\
58108.65545  &       -1.501  &  1.233 &    1.312  &   180.5   &  -94.5   &  84.2  &            &  \\
58109.54377  &       -1.485  &  1.240 &    1.314  &   177.0   &  -87.3   &  87.2  &            &  \\
58110.54374  &       -1.417  &  1.246 &    1.316  &   174.6   &  -90.6   &  80.2  &            &  \\
58111.56345  &       -1.406  &  1.214 &    1.315  &   174.1   &  -92.9   &  80.1  &            &  \\
58112.56647  &       -1.417  &  1.247 &    1.332  &   181.2   &  -94.4   &  85.8  &            &  \\
58122.57100  &       -1.180  &  1.199 &    1.292  &   181.0   &  -105.4  &  75.1  &            &  \\
58122.67347  &       -1.165  &  1.197 &    1.280  &   183.2   &  -107.5  &  73.5  &            &  \\
58122.77185  &       -1.120  &  1.198 &    1.280  &   184.3   &  -105.2  &  75.2  &            &  \\
58123.56582  &       -1.129  &  1.199 &    1.276  &   178.9   &  -100.1  &  76.1  &            &  \\
  \\			    
 \hline 		    				     
 \end{tabular}  	    					      
 \label{tab.HeI}	    			    
 \end{table*}		    			    
%--------------------------  ----------------
\begin{table*}
  \centering
  \addtocounter{table}{-1}
  \caption{ continuation. 
  }
   \begin{tabular}{lrrrrrrrcc} 
   \hline
   \\
58124.59271  &       -1.115  &  1.198 &    1.265  &   173.3   &  -98.1   &  74.1  &            &  \\
58124.68994  &       -1.113  &  1.197 &    1.273  &   170.8   &  -98.3   &  73.4  &            &  \\
58124.78338  &       -1.134  &  1.200 &    1.281  &   172.2   &  -97.6   &  71.7  &            &  \\
58125.56759  &       -1.114  &  1.190 &    1.277  &   173.0   &  -97.2   &  74.3  &            &  \\
58125.66887  &       -1.176  &  1.190 &    1.281  &   175.5   &  -97.3   &  73.8  &            &  \\
58125.76956  &       -1.124  &  1.191 &    1.268  &   178.0   &  -99.8   &  73.1  &            &  \\
58319.96541  &       -0.687  &  1.074 &    1.149  &   209.9   &  -156.8  &  35.3  &            &  \\
58332.97367  &       -0.707  &  1.078 &    1.171  &   219.8   &  -182.9  &  37.3  &            &  \\
58348.92388  &       -0.857  &  1.105 &    1.186  &   210.5   &  -178.2  &  32.3  &            &  \\
58366.87575  &       -0.936  &  1.108 &    1.197  &   203.8   &  -176.2  &  35.9  &            &  \\
58386.87432  &       -1.162  &  1.141 &    1.198  &   233.5   &  -204.5  &  35.7  &            &  \\
58401.81034  &       -1.269  &  1.166 &    1.210  &   215.9   &  -184.4  &  36.9  &            &  \\
58446.76209  &       -1.383  &  1.207 &    1.197  &   217.8   &  -183.2  &  32.8  &            &  \\
58458.81081  &       -1.319  &  1.159 &    1.193  &   223.6   &  -192.5  &  32.1  &            &  \\
58470.80746  &       -1.326  &  1.170 &    1.196  &   223.3   &  -185.8  &  35.0  &            &  \\
58482.81011  &       -1.475  &  1.196 &    1.204  &   211.6   &  -179.6  &  30.0  &    201.7   &  \\
58494.81083  &       -1.394  &  1.187 &    1.202  &   201.5   &  -177.1  &  24.7  &    174.4   &  \\
58507.55874  &       -1.454  &  1.218 &    1.171  &   206.1   &  -174.9  &  33.9  &    208.6   &  \\
58521.57478  &       -1.381  &  1.188 &    1.167  &   196.6   &  -167.8  &  28.4  &    217.5   &  \\
58544.58365  &       -1.351  &  1.206 &    1.167  &   179.0   &  -142.5  &  36.9  &    232.3   &  \\
58682.52460  &       -0.896  &  1.223 &    1.089  &   367.4   &  -117.1  &  242.3 &            &  \\
58683.53920  &       -0.873  &  1.229 &    1.081  &   357.9   &  -120.7  &  235.8 &            &  \\
58683.54370  &       -0.817  &  1.228 &    1.106  &   358.4   &  -116.9  &  244.8 &            &  \\
58714.53390  &       -0.813  &  1.180 &    1.088  &   348.2   &  -104.4  &  242.5 &            &  \\
58714.54510  &       -0.781  &  1.176 &    1.083  &   353.4   &  -107.6  &  244.7 &            &  \\
58718.46660  &       -0.796  &  1.182 &    1.081  &   345.4   &  -104.5  &  239.8 &            &  \\
58718.47190  &       -0.808  &  1.192 &    1.104  &   341.3   &  -103.1  &  240.1 &            &  \\
58741.53080  &       -0.716  &  1.197 &    1.081  &   331.9   &  -95.1   &  240.6 &            &  \\
58741.53800  &       -0.727  &  1.191 &    1.091  &   335.8   &  -92.9   &  235.8 &            &  \\
58803.45210  &       -0.586  &  1.129 &    1.092  &   279.2   &  -72.6   &  208.5 &            &  \\
58803.46410  &       -0.570  &  1.120 &    1.091  &   271.5   &  -72.6   &  201.9 &            &  \\
58823.35790  &       -0.510  &  1.115 &    1.080  &   257.3   &  -67.3   &  195.5 &            &  \\
58823.36280  &       -0.518  &  1.122 &    1.091  &   260.1   &  -67.6   &  197.3 &            &  \\
58824.29680  &       -0.521  &  1.118 &    1.092  &   263.3   &  -73.2   &  192.4 &            &  \\
58824.30150  &       -0.494  &  1.121 &    1.094  &   267.6   &  -74.4   &  192.9 &            &  \\
58852.25040  &       -0.436  &  1.110 &    1.069  &   244.1   &  -67.8   &  179.2 &            &  \\
58852.25530  &       -0.446  &  1.118 &    1.079  &   254.5   &  -68.4   &  179.4 &            &  \\
58856.20920  &       -0.422  &  1.107 &    1.081  &   225.4   &  -67.6   &  160.1 &            &  \\
58856.23810  &       -0.413  &  1.088 &    1.087  &   231.4   &  -67.2   &  163.2 &            &  \\
58859.22110  &       -0.401  &   ---  &     ---   & 	---   &   ---	 &   ---  & 	       &  \\
58859.23000  &       -0.417  &  1.098 &    1.073  &   223.4   &  -70.8   &  152.7 &            &  \\
58865.33290  &       -0.417  &  1.086 &    1.078  &   216.4   &  -64.9   &  150.5 &            &  \\
  \\	    		      									 
 \hline     						        				 
 \end{tabular}										      	 
 \end{table*} 				      							 
%-------------------------------------- 							 

\bsp	% t  yp  esetting comme  nt								 
\label{lastpage}										 
	    											 
\end{document}